\documentclass[journal]{new-aiaa} 

\usepackage[utf8]{inputenc}
\usepackage{textcomp}

\usepackage{graphicx}
\usepackage{amsmath}
\usepackage[version=4]{mhchem}
\usepackage{siunitx}
\usepackage{longtable,tabularx}
\usepackage{subfigure}
\usepackage[export]{adjustbox}
\usepackage{epstopdf}

\title{First-Principle-Inspired Reduced-Order Models of Chemical-Kinetics in $\text{H}_2\left(\text{X}^1\Sigma_g^+\right)$+$\text{H}\left({}^2\text{S}\right)$ System}

\author{Hye Su Jeong \footnote{Graduate Student, Department of Aerospace Engineering.} and Tae Woong Jeong\footnote{Graduate Student, Department of Aerospace Engineering.}}
\affil{Korea Advanced Institute of Science and Technology, Daejeon 34141, Republic of Korea}
\author{Sung Min Jo\footnote{Assistant Professor, Department of Mechanical and Aerospace Engineering, sungmin.jo@ucf.edu (Corresponding Author).}}
\affil{UCF Center of Excellence in Hypersonics and Space Propulsion (HYPERSPACE), \\ University of Central Florida, Orlando, FL 32816, USA}

\begin{document}

\maketitle

\begin{abstract}
In the present study, two-different reduced-order models are proposed for $\text{H}_2\left(\text{X}^1\Sigma_g^+\right)$+$\text{H}\left({}^2\text{S}\right)$ system by leveraging first-principle quasi-classical trajectory simulations and in-depth master equation analyses. The most recent available \emph{ab-initio} potential energy surface is adopted to construct a new set of rovibrational state-to-state kinetic database valid over a wide range of temperatures. Firstly, a modified two-temperature model is proposed by incorporating the master equation-informed model parameters, enabling the advanced treatment of the internal energy coupling and the nonequilibrium dissociation predictions. Secondly, a hybrid coarse-graining model is proposed by combining a graph-based approach optimized globally for a wide range of temperatures with a centrifugal-barrier-based coarse-graining method. The proposed reduced-order models offer significantly improved accuracy in predicting the nonequilibrium energy transfer and dissociation dynamics compared to the existing coarse-graining and 2T models in previous studies. In addition, aerothermal heating prediction relevant to Uranus planetary entry reveals 16.5\% of convective heat flux discrepancy compared to the present modified 2T approach with the existing 2T, demonstrating the importance of accurate modeling of the chemical-kinetics in the $\text{H}_2\left(\text{X}^1\Sigma_g^+\right)$+$\text{H}\left({}^2\text{S}\right)$ system.
\end{abstract}

\section{\label{sec:1}Introduction}
\lettrine{G}{iant} planets have emerged as key targets for future space exploration missions. The Decadal Survey 2023–2032~\cite{NAP26522} prioritizes the Uranus Orbiter and Probe mission and emphasizes the importance of giant planet studies. The giant planets Jupiter, Saturn, Uranus, and Neptune have atmospheres that mainly consist of diatomic hydrogen (\ce{H2}) with small amounts of helium (\ce{He}). Therefore, it is essential to understand thermochemical nonequilibrium flow characteristics of the hydrogen-dominated atmospheres during hypersonic entry of probes to design a reliable thermal protection system (TPS). During hypersonic atmospheric entry, kinetic energy is converted into internal energy across strong shock waves. As a result, the internal energy modes of gas particles become excited and then relax toward thermochemical equilibrium, while chemical-kinetic processes such as dissociation and ionization occur. A proper understanding of such chemical-kinetics has an important impact on aerothermal heating characterization~\cite{jo2019electronic}. For the giant planetary probe, a three-body interaction, $\text{H}_2$+H, is one of the key collisional processes impacting the aerothermal environment \cite{palmer2014aeroheating}.
Gradients in temperature and species concentrations developed during the collisional process significantly influence aerothermal heating at the TPS surface.
According to Computational Fluid Dynamics (CFD) studies of hypersonic atmospheric entry to the giant planets~\cite{santos2019computational, palmer2014aeroheating}, the maximum stagnation-line temperature can reach approximately 28{,}000~K. 
Hence, a high-fidelity chemical-kinetics model capable of accurate prediction of these nonequilibrium effects is imperative.

In several flow simulation studies~\cite{palmer2014aeroheating,higdon2018direct,coelho2023aerothermodynamic}, hydrogen dissociation rate coefficients proposed by Leibowitz \textit{et al.}~\cite{leibowitz1972measurements,leibowitz1976ionizational} have been widely used. The rate coefficients of Leibowitz \textit{et al.}~\cite{leibowitz1972measurements} are based on the decomposition rate measurements of Jacobs \textit{et al.}~\cite{jacobs1967kinetics}, which were conducted at temperatures below 5{,}000~K. Leibowitz \textit{et al.}~\cite{leibowitz1976ionizational} described results from a comparison between shock-tube measurements and numerical solutions of reaction kinetic equations, but the origin and temperature range of the underlying data were not clearly specified.
Cohen~\textit{et al.}~\cite{cohen1983chemical} reviewed measured dissociation rate coefficients for the hydrogen chemistry, but their data was limited to temperatures below 5{,}000~K. Given that the temperature range encountered during hypersonic atmospheric entry to the giant planets may exceed the coverage of the experimental data in the review, computational approaches are required to fill the knowledge gap. 

For efficient constructions of accurate chemical-kinetic databases of the hydrogen, 
quasi-classical trajectory (QCT) methods~\cite{jaffe2015first, Truhlar1979} over \textit{ab-initio} or semi-empirical potential energy surfaces (PESs) have been employed~\cite{schwenke1990theoretical, martin1996master, ESPOSITO1999636, kim2009master, vargas2024state}. 
Since QCT calculations are highly sensitive to the accuracy of the PES, it is important to use the most accurate PES available. In previous studies, several attempts have been made to perform QCT calculations for the $\text{H}_2\left(\text{X}^1\Sigma_g^+\right)$+$\text{H}\left({}^2\text{S}\right)$ system using \textit{ab-initio} PESs. Martin~\textit{et al.}~\cite{martin1996master} and Esposito~\textit{et al.}~\cite{ESPOSITO1999636} used the LSTH PES~\cite{siegbahn1978accurate} developed by Siegbahn and Liu. Kim~\textit{et al.}~\cite{kim2009master} used the BKMP2 PES~\cite{boothroyd1996refined} constructed by Boothroyd \emph{et al.} to perform the QCT and master equation analysis. 
Although the LSTH and BKMP2 PESs are in \textit{ab-initio} accuracy, they are limited by the use of non–full configuration interaction methods. In addition, the LSTH and BKMP2 PES share many data points in common. The CCI PES~\cite{mielke2002hierarchical} was fitted to newly calculated energies obtained using a near full configuration interaction method with extrapolation to the complete basis set. Based on the CCI PES, Mielke \textit{et al.}~\cite{mielke2009functional} introduced the BH PES in 2009, which includes the Born–Oppenheimer Diagonal Correction (BODC). This correction accounts for the interaction between nuclear motion and electronic states neglected in the Born–Oppenheimer approximation. In the present study, therefore, the BH PES is used for the QCT calculations.
Once the state-to-state (StS) cross sections and rate coefficients describing internal energy excitation and dissociation are obtained through QCT simulations, the master equation analysis can be performed to characterize energy transfer and chemical reactions among internal states in detail.
Previous studies~\cite{martin1996master,ESPOSITO1999636} investigated global dissociation rate coefficients using master equation analyses but did not analyze nonequilibrium relaxation behavior, which is essential for hypersonic CFD simulations. In particular, Vargas \textit{et al.}~\cite{vargas2024state} performed QCT calculations using the CCI PES without master equation analysis. Kim \textit{et al.}~\cite{kim2009master} examined the evolution of rovibrational energy states using a master equation approach but did not propose a reduced-order model that can be directly applied to hypersonic CFD simulations.

In addition to the QCT calculations, recent work by Carroll \textit{et al.}~\cite{carroll2025treatment} modeled nonequilibrium dissociation effects of \ce{H2} based on existing experimental and computational data, and proposed practical rate coefficient formulations that can be directly used in macroscopic CFD analyses. By distinguishing thermal equilibrium, quasi-steady-state (QSS), and pre-QSS regimes, they showed that in the QSS regime the chemical source term can be expressed using only the steady dissociation rate without recombination. They also demonstrated that this formulation reproduces the results of 0-D reactor master equation calculations when the third body is \ce{H2}, \ce{H}, or \ce{He}. However, since their study relied on previously available data, the need remains for more up-to-date \textit{ab-initio} full StS-based studies and for a detailed investigation of energy transfer mechanisms in addition to dissociation.
In the recent study by Colonna \textit{et al.} \cite{colonna2026multi}, a vibration-specific StS approach, applicable to weakly ionized plasmas with an electron energy distribution, is used to develop and demonstrate a multi-temperature (mT) model for hydrogen/He mixtures. However, the rotational contribution, which can be important for the nonequilibrium chemical-kinetics of hydrogen, was neglected in the model investigation.

Moreover, directly applying \textit{ab-initio} full StS approaches to CFD simulations remains computationally impractical.
In the present work, $\text{H}_2\left(\text{X}^1\Sigma_g^+\right)$ has 343 rovibrational states, resulting in more than \(10^{5}\) StS rate coefficients for physico-chemical processes such as energy transfer and dissociation, therefore, advanced reduced-order modeling is investigated. In this vein, the existing QCT calculations and master equation analysis studies for the hydrogen systems~\cite{schwenke1990theoretical, martin1996master, ESPOSITO1999636, kim2009master, vargas2024state} have not proposed such reduced-order representations that can be applicable to the macroscopic flow calculations while derived from \textit{ab-initio} quantum chemistry method on the state-of-the-art PES. In this regard, the present study proposes a full StS hydrogen chemistry model for the $\text{H}_2\left(\text{X}^1\Sigma_g^+\right)$+$\text{H}\left({}^2\text{S}\right)$ collisional interaction based on QCT calculations using the most recent and accurate PES~\cite{mielke2009functional}. Then, to enable the application of the characterized chemical-kinetics obtained from the master equation study to hypersonic CFD simulations, reduced-order models are proposed in two distinct forms; a modified version of the conventional two-temperature (2T) model~\cite{park1989nonequilibrium} and a physics-informed advanced coarse-graining representation, which can be extended for four-body interactions \cite{macdonald2018construction,jo2023rovibrational}, for example, $\text{H}_2\left(\text{X}^1\Sigma_g^+\right)$+$\text{H}_2\left(\text{X}^1\Sigma_g^+\right)$.

The remaining part of the paper is organized as follows.
In Sec.~\ref{sec:physical_modeling}, the rovibrational StS chemical-kinetic database constructed using a QCT method is described, followed by the details of the master equation formulations and the derivations of the proposed reduced-order models. Section~\ref{sec:4} presents the results of in-depth master equation analysis for canonical test cases for characterizing the physical processes and for comparing various models' performance. In addition, multi-dimensional hypersonic flow simulations are carried out to compare the proposed modified 2T model with the existing 2T approach in the literature for a trajectory point of hypersonic atmospheric entry to Uranus. 
Section~\ref{sec:5} summarizes the conclusions of the present work.

\section{\label{sec:physical_modeling}Physical Modeling}

\subsection{\label{sec:database}Rovibrational State-to-State Kinetic Database}

In this work, rovibrational-specific StS kinetic database has been constructed by means of QCT simulations using \textsc{CoarseAIR} software package~\cite{venturi2020bayesian,venturi2020data}, which is a modernized version of NASA’s \textsc{vvtc} toolbox~\cite{schwenke1988calculations}. The latest BH PES~\cite{mielke2009functional} for the $\text{H}_2\left(\text{X}^1\Sigma_g^+\right)$+$\text{H}\left({}^2\text{S}\right)$ system has been implemented to \textsc{CoarseAIR}. In total, the 343 rovibrational levels with the maximum vibrational quantum number, \(v_{\text{max}}{=}14\), including both bound and quasi-bound states were determined by solving Schr$\ddot{\text{o}}$dinger's equation based on the WKB semi-classical approximation~\cite{Truhlar1979,schwenke1988calculations}.
The QCT calculations were carried out for relative translational energies ranging from 0.1 eV to 30 eV, which covers the high-energy collisions prevalent in hypersonic flow regimes. For each energy, 40,000 trajectories were used per collision pair to secure statistical convergence at the order of $10^{-2}$ for the ratio between the cross section and its standard deviation. The impact parameter, $b$, was treated by a stratified sampling method with a discretized grid of \(b_{\text{max}}=[0.25, 0.5, 1, 2.5, 4, 7, 12] \, a_{0}\). Once the StS cross sections are obtained, the corresponding state-resolved bound-bound and bound-free rate coefficients can be calculated as: 

\begin{equation}
\label{eq:k_i_to_j}
k_{i\to j}(T)
= \sqrt{\frac{8 k_{\mathrm{B}} T}{\pi \mu}}
  \int_{0}^{E_{\text{rel},\max}} \sigma_{i\to j}(E_{\text{rel}})\, E_{\text{rel}} \,
  \exp\!\left(-\frac{E_{\text{rel}} }{k_{\mathrm{B}} T}\right)\, dE_{\text{rel}} ,
\end{equation}

\begin{equation}
\label{eq:k_i_to_c}
k_{i\to c}(T)
= \sqrt{\frac{8 k_{\mathrm{B}} T}{\pi \mu}}
  \int_{0}^{E_{\text{rel},\max}} \sigma_{i\to c}(E_{\text{rel}})\, E_{\text{rel}} \,
  \exp\!\left(-\frac{E_{\text{rel}}}{k_{\mathrm{B}} T}\right)\, dE_{\text{rel}} .
\end{equation}

\noindent
where the indices \(i\) and \(j\) denote rovibrational states, and \(c\) implies the dissociated state. Each rovibrational state $i$ corresponds to a unique pair of vibrational \((v)\) and rotational \((J)\) quantum number (\emph{i.e.}, $i{=}(v,J)$). \(\sigma\) and \(k\) respectively denote the cross section and the rate coefficient. \(k_{\mathrm{B}}\) is the Boltzmann constant, and \(\mu\) is the reduced mass. \(T\) represents the translational temperature, and $E_{\text{rel}}$ is the relative translational energy of the collision pair. The rate coefficients obtained from Eqs.~(\ref{eq:k_i_to_j}) and (\ref{eq:k_i_to_c}) are respectively for the collisional excitation/de-excitation (Eq.~(\ref{eq:rovib_transfer})) and dissociation/recombination (Eq.~(\ref{eq:diss_recomb})) processes:

\begin{equation}
\label{eq:rovib_transfer}
\ce{H2(\mathit{i}) + H <=>[$k_{\mathit{i}\to \mathit{j}}$][$k_{\mathit{j}\to \mathit{i}}$] H2(\mathit{j}) + H} ,
\end{equation}

\begin{equation}
\label{eq:diss_recomb}
\ce{H2(\mathit{i}) + H <=>[$k_{\mathit{i}\to \mathit{c}}$][$k_{\mathit{c}\to \mathit{i}}$] H + H + H}.
\end{equation}

In the present study, the exothermic bound-bound transition (\emph{i.e.}, de-excitation) rate coefficients were obtained from the QCT calculations for the inelastic and exchange processes to ensure favorable statistical quality. In addition, the present QCT calculations model the dissociation process rather than the recombination. The excitation and the recombination rate coefficients are then reconstructed by imposing the micro-reversibility~\cite{park1989nonequilibrium} over the opposite counterparts to guarantee proper equilibration of the chemical system.
All the physical processes were assumed to occur on the electronic ground states of the molecular and atomic hydrogen by neglecting non-adiabatic processes. 

\begin{figure}[hbt!]
    \centering
    \subfigure[Cross sections for H$_2$($v$=0, $J$=0)+H{$\rightarrow$}H$_2$($v$=1, $J$=1)+H.]
    {
        \includegraphics[width=0.45\textwidth]{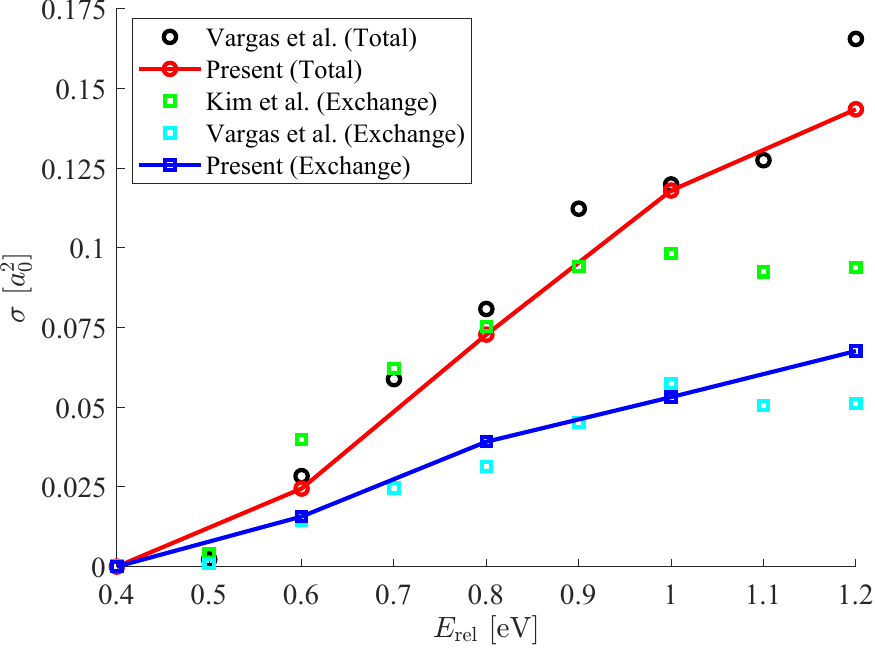}
        \label{fig:CS}
    }
    \subfigure[Thermal dissociation rate coefficients.]
    {
        \includegraphics[width=0.45\textwidth]{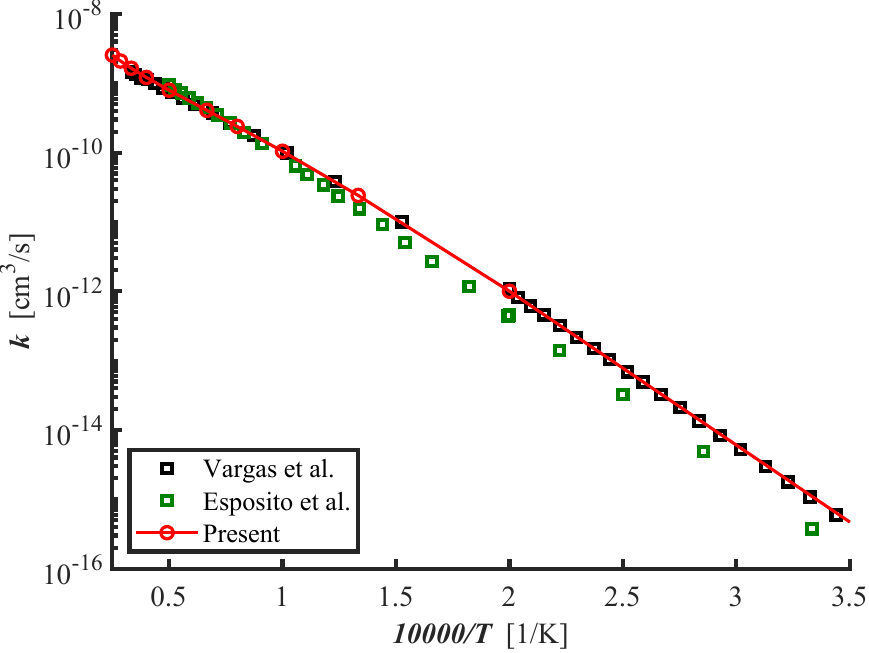}
        \label{thermal_dissociation_rate}
    }
  \caption{Comparisons of the (a) bound-bound transition cross sections and (b) thermal dissociation rate coefficients with the literature data \cite{kim2009master,vargas2024state,ESPOSITO1999636}.}
    \label{Validation}
\end{figure}

The present QCT calculations are then validated by comparing with the available computational results in the literature, as shown in Fig. \ref{Validation}.
Figure~\ref{fig:CS} compares the rovibrational StS cross sections for the $\text{H}_2(v=0,J=0)+\text{H} \rightarrow \text{H}_2(v=1,J=1)+\text{H}$ transition obtained in this work with those reported by the previous studies \cite{kim2009master,vargas2024state}.
To the best of the authors’ knowledge, no previous literature has reported cross sections obtained using the BH PES. However, since the BH PES was developed on top of the CCI PES with modifications to account for the mass dependence of hydrogen, it is expected to yield a similar trend in the cross section profile. The present exchange and inelastic cross sections show reasonable agreement with the results of Vargas \emph{et al.}~\cite{vargas2024state} obtained using the CCI PES. On the other hand, the discrepancy between the present result and the data from Kim~\textit{et al.}~\cite{kim2009master} above 0.9 eV can be attributed to the use of different PESs. Figure \ref{thermal_dissociation_rate} compares the present thermal dissociation rate coefficients with the QCT-based literature data \cite{vargas2024state,ESPOSITO1999636}. The three-different thermal dissociation rate coefficients are in reasonable agreement, although a certain degree of discrepancy is observed against the result of Esposito \textit{et al.}~\cite{ESPOSITO1999636} in the low-temperature regime, attributed to the use of the different PESs. 

\subsection{\label{sec:methods_StS}Rovibrational State-to-State Master Equations}

The rovibrational StS kinetic database explained in Sec.~\ref{sec:database} was then employed for the master equation study of the $\text{H}_2\left(\text{X}^1\Sigma_g^+\right)$+$\text{H}\left({}^2\text{S}\right)$ system. Equation~(\ref{eq:i_ME}) describes the time rate of change of $i$-th rovibrational state $n_i$, and Eq.~(\ref{eq:H_ME}) denotes the time rate of change of atomic hydrogen number density $n_{\text{H}}$ due to the chemical dissociation of $\text{H}_2$:

\begin{equation}
\label{eq:i_ME}
\frac{dn_{i}}{dt}
= \sum_{j} \left( k_{j \to i}\, n_{j}\, n_{\mathrm{H}}- k_{i \to j}\, n_{i}\, n_{\mathrm{H}} \right)+ k_{c \to i}\, n_{\mathrm{H}}^{3}- k_{i \to c}\, n_{i}\, n_{\mathrm{H}} ,
\end{equation}

\begin{equation}
\label{eq:H_ME}
\frac{dn_{\mathrm{H}}}{dt}
= -\frac{dn_{\ce{H2}}}{dt}
= 2 \sum_{i} \left( k_{i \to c}\, n_{i}\, n_{\mathrm{H}}- k_{c \to i}\, n_{\mathrm{H}}^{3} \right).
\end{equation}

\noindent
Equation~(\ref{eq:i_ME}) is defined for individual rovibrational states; thus, a total of 344 ordinary differential equations are solved in the present study. The master equations are then numerically integrated using \textsc{plato} (PLAsmas in Thermodynamic nonequilibrium) library~\cite{munafo2020computational,munafo2025plato} for idealized 0-D isochoric and isothermal heat-bath conditions. The initial pressure and temperature were set to 1{,}000~Pa and 300~K, respectively, with the initial number densities of $n_{\text{H}_2}=n_{\text{H}}=1.2072\times10^{23}$ cm${}^{-3}$. Then the translational temperature of the system is instantaneously scaled to higher values, ranging from 1{,}000~K to 50{,}000~K to characterize the internal energy transfer and chemical dissociation mechanisms.

With the solutions of the rovibrational-specific master equations (\emph{i.e.}, $n_i$ and $n_{\text{H}}$), the average nonequilibrium energy can be computed as:
\begin{equation}
E_{X}=\frac{\sum_{i} e_{x}(i)\, n_{i}} {\sum_{i} n_{i}},
\label{eq:Average_Energy}
\end{equation}

\noindent 
where the subscripts $X$ and $x$ respectively imply the average and the state-specific contributions of the nonequilibrium energy modes, such as rotational ($X{=}R$ and $x{=}r$), vibrational ($X{=}V$ and $x{=}v$), and internal ($X{=}I$ and $x{=}\text{int}$) components. In the present study, the vibrational and rotational term energies of $i$-th state are correspondingly defined as $e_v(i)=e_{\text{int}}(v,J=0)$ and $e_r(i)=e_{\text{int}}(v,J)-e_v(i)$ where $i=(v,J)$ to allow separation of the rotational and vibrational contributions. Using the average nonequilibrium energy in Eq. (\ref{eq:Average_Energy}), the corresponding temperatures can be computed as follows:
\begin{equation}
    \label{eq:T_R}
    E_R - E^B_R\left(T_R, T_V\right)=0,
\end{equation}
\begin{equation}
    \label{eq:T_V}
    E_V - E^B_V\left(T_R, T_V\right)=0,
\end{equation}
\begin{equation}
    \label{eq:T_int}
    E_I - E_I^B\left(T_I\right)=0,
\end{equation}

\noindent
where $E_X^B$ for each energy mode denotes the average energy based on a Maxwell-Boltzmann distribution of the internal states defined by the partition function. The temperatures, $T_R$, $T_V$, and $T_I$, can be evaluated by solving Eqs. (\ref{eq:T_R})--(\ref{eq:T_int}) using a Newton-Raphson method. In addition, the population-averaged global dissociation rate coefficients during the QSS period of $\text{H}_2$ can be defined as: 

\begin{equation}
\label{eq:k_QSS}
k_{\mathrm{QSS}} =
\frac{\sum_{i} n_{i}^{\mathrm{QSS}} \, k_{i \to c}}
     {\sum_{i} n_{i}^{\mathrm{QSS}}} .
\end{equation}

\noindent
During the dissociation or recombination of $\text{H}_2$, the internal energy is either lost or gained, for example, the rate of vibrational energy loss $\dot{e}_v^{\mathrm{loss}}$ can be expressed as:

\begin{equation}
\dot{e}_v^{\mathrm{loss}}
= \sum_{i} e_v(i)\left(k_{i\to c}\, n_i\, n_{\mathrm{H}}-k_{c\to i}\, n_{\mathrm{H}}^{3}\right).
\label{eq:Ev_loss_rate}
\end{equation}

\noindent
Therefore, the vibrational energy loss ratio $C^{DV}$ can be defined as: 

\begin{equation}
C^{DV}
= \frac{1}{D_{\text{H}_2}} \frac{\sum_{i} e_v(i)\left(k_{i\to c}\, n_i\, n_{\mathrm{H}}-k_{c\to i}\, n_{\mathrm{H}}^{3}\right)}
{\sum_{i}\left(k_{i\to c}\, n_i\, n_{\mathrm{H}}-k_{c\to i}\, n_{\mathrm{H}}^{3}\right)} .
\label{eq:epsilon_bar_v}
\end{equation}

\noindent
where $D_{\text{H}_2}$ denotes the average dissociation potential of $\text{H}_2$, 4.518 eV. Similarly, the rotational and internal energy loss ratios can be calculated by replacing the state vibrational energy $e_v(i)$ in Eq. (\ref{eq:epsilon_bar_v}) with $e_r(i)$ and $e_{\text{int}}(i)$, respectively.

\subsection{\label{sec:methods_ROM}Reduced-Order Modeling}
\subsubsection{\label{sec:level3}Modified Two-Temperature Model}

By leveraging the results of the rovibrational StS master equation analysis, we propose a new modified 2T model for the nonequilibrium chemical-kinetics of $\text{H}_2\left(\text{X}^1\Sigma_g^+\right)$+$\text{H}\left({}^2\text{S}\right)$ system. In a 2T model, the thermal energy modes are separated into the translational–rotational (TR) and the vibrational (V) energy modes when the influences of ionization and excited electronic state are neglected~\cite{park1989nonequilibrium}. Under the assumption of the electronic ground state, the vibrational energy conservation equation reads:

\begin{equation}
\frac{\partial}{\partial t} \rho e_{v} +\frac{\partial}{\partial x^j}\left[\left(\rho e_{v}\right) u^j\right]=\frac{\partial}{\partial x^j}\left(\eta_{v} \frac{\partial T_{V}}{\partial x^j}\right) +\frac{\partial}{\partial x^j}\left(\rho \sum_{s=1}^{a l l} h_{e v, s} D_s \frac{\partial y_s}{\partial x^j}\right)+ \Omega_{V-T} + \Omega_{c-V},
\label{eq:v}
\end{equation}

\noindent
where $\rho$ is density, and $x^j$ and $u^j$ are the position and velocity vectors of the $j$ coordinate. $e_v$ and $\eta_v$ are correspondingly the specific energy and thermal conductivity of the vibrational energy mode. $h_{ev,s}$, $D_s$, and $y_s$ are the specific enthalpy of the electronic ground and the vibrational energy modes, diffusion coefficient, and mass fraction of species $s$, respectively. $\Omega_{V-T}$ is the rate of internal energy transfer in the vibrational-to-translational (V-T) energies. In the present study, $\Omega_{V\text{-}T}$ is modeled using the Landau--Teller formulation \cite{LT_1936}:.

\begin{equation}
\Omega_{V\text{-}T} = \rho \frac{e^*_{v} - e_{v}}{\tau_{VT}},
\label{eq:omega_vT}
\end{equation}

\noindent
where $e_v^*$ is the specific vibrational equilibrium energy, and $\tau_{VT}$ is the vibrational relaxation time of $\text{H}_2$, respectively. In the present study, $\tau_{VT}$ is determined using the $e$-folding method over results of the rovibrational StS master equation analysis, including the high-temperature correction \cite{park1989nonequilibrium} that reads:

\begin{equation}
\label{eq:tau_total}
\tau_{VT} = \tau_{MW} + \tau_{c},
\end{equation}

\begin{equation}
\label{eq:tau_MW}
\tau_{MW} = 
\frac{101325}{p}
\exp\!\left[A_{VT}\!\left(T^{-1/3} - B_{VT}\right) - 18.42\right],
\end{equation}

\begin{equation}
\label{eq:tau_c}
\tau_{c} =
\left(
  n_{\text{H}_2}\sigma
  \sqrt{\frac{8k_{\mathrm{B}}T}{\pi \mu}}
\right)^{-1},
\end{equation}

\begin{equation}
\label{eq:sigma}
\sigma = C_{VT} T^{D_{VT}},
\end{equation}

\noindent
where $\tau_{MW}$ is described using the Millikan-White (MW) expression \cite{millikan1963systematics} and $\tau_c$ is the high-temperature correction \cite{park1989nonequilibrium} described by the collision limiting cross section $\sigma$. In the previous hypersonic flow simulations of hydrogen systems \cite{palmer2014aeroheating,higdon2018direct,coelho2023aerothermodynamic,santos2019computational}, the high-temperature correction has been neglected. In the present study, the parameters $A_{VT}$, $B_{VT}$, $C_{VT}$, and $D_{VT}$ are determined by leveraging the rovibrational StS master equation results over a wide range of temperature to enable the use of those in hypersonic CFD calculations.

$\Omega_{c-V}$ is the net vibrational energy loss due to the dissociation, which is often noted as a chemistry-vibrational (c-V) energy coupling term in the literature \cite{gnoffo1989conservation}. In the present study, a preferential dissociation model is considered to take into account the nonuniform probability of the dissociation over the molecular diatomic potential that reads:

\begin{equation}
\Omega_{c\text{-}V} = C^{DV} \dot{\omega}_{\text{H}_2}^D D_{\text{H}_2},
\label{eq:omega_cv}
\end{equation}

\noindent
where $\dot{\omega}_{\text{H}_2}^D$ is the mass rate of production of $\text{H}_2$ due to the dissociation. The discrete values of $C^{DV}$ determined from the rovibrational StS master equation analysis using Eq. (\ref{eq:epsilon_bar_v}) are parameterized as a continuous function of temperature:

\begin{equation}
\label{eq:cd-fit}
C^{DV} =
\exp\!\left(
  \frac{K_{1}^V}{T} + K_{2}^V + K_{3}^V\ln T + K_{4}^V T + K_{5}^V T^{2}
\right) .
\end{equation}

\noindent
In calculating $\Omega_{c-V}$ by means of the modified 2T model of the present study, Eq. (\ref{eq:cd-fit}) is used to accommodate continuous derivatives over the temperature.  

In the 2T model, the species mass conservation equations are constructed as follows:

\begin{equation}
\frac{\partial}{\partial t} \rho_s+\frac{\partial}{\partial x^j} \rho_s u^j=\frac{\partial}{\partial x^j}\left(\rho D_s \frac{\partial}{\partial x^j} y_s\right)+\dot{\omega}_s^D,
\end{equation}

\noindent
where $\rho_s$ is partial density of species $s$ (\emph{i.e.}, H or $\text{H}_2$), and $\dot{\omega}_s^D$ can be defined as:

\begin{equation}
\dot{\omega}_{s}^D = M_{s}\left(\beta_{s} - \alpha_{s}\right)\left(R_{f}-R_{b}\right),
\label{eq:omega_s}
\end{equation}

\noindent
where $M_s$ is the molar mass of species $s$. $\alpha_s$ and $\beta_s$ are the species-wise stoichiometric coefficients of the reactants and the products of the dissociation reaction in Eq. (\ref{eq:diss_recomb}). The forward and reverse reaction rates, $R_f$ and $R_b$, are defined by:

\begin{equation}
R_{f}=1000\left[k_{f}\prod_{s=1}^{\mathrm{all}}\left(0.001\,\rho_{s}/M_{s}\right)^{\alpha_{s}}\right],
\label{eq:R_fk}
\end{equation}

\begin{equation}
R_{b}=1000\left[k_{b}\prod_{s=1}^{\mathrm{all}}\left(0.001\,\rho_{s}/M_{s}\right)^{\beta_{s}}\right].
\label{eq:R_bk}
\end{equation}

\noindent
Here, $k_{f}$ and $k_{b}$ denote the forward (\emph{i.e.}, dissociation) and backward (\emph{i.e.}, recombination) rate coefficients, respectively. The factors 0.001 and 1000 are unit conversion factors~\cite{kim2021modification}. 
The forward rate coefficient $k_{f}$ is obtained from the rovibrational StS master equation analysis, in particular, during the QSS period of $\text{H}_2$ using Eq. (\ref{eq:k_QSS}). The backward rate coefficient $k_b$ is then calculated based on the micro-reversibility with respect to the forward reaction. To facilitate the use of the forward rate coefficient in the modified 2T model, a modified Arrhenius type of equation is used as:

\begin{equation}
\label{eq:k-fitting}
k_f(T_a) = A \, T_a^{n} \exp\!\left(-\frac{C}{T_a}\right),
\end{equation}

\noindent
here $T_a$ is rate control temperature \cite{park1989nonequilibrium} defined as $T_a=\sqrt{T_{TR} T_V}$ where $T_{TR}$ is the translational-rotational temperature to take into account the rotational-vibrational-translational (RVT) energy transfers. The result of Eq. (\ref{eq:k_QSS}) is directly used to determine the parameters, $A$, $n$, and $C$ of Eq. (\ref{eq:k-fitting}).
As a result, the present modified 2T model for the $\text{H}_2\left(\text{X}^1\Sigma_g^+\right)$+$\text{H}\left({}^2\text{S}\right)$ system consists of the following updated model parameters:
(1) $\tau_{VT}$,
(2) $C^{DV}$, and
(3) $k_f(T_a)$.

\subsubsection{\label{sec:method_HB}Hybrid Coarse-Graining}

The second reduced-order model we propose is a Hybrid Coarse-Graining (HCG), a physics-based, optimized representation of conventional coarse-graining strategies \cite{liu2015general,macdonald2018construction,sahai2017adaptive,venturi2020data,sharma2020coarse,Notey2025,jo2023rovibrational,jo2022rovibrational}. In a coarse-graining model, rovibrational energy states are grouped together to reduce number of unknowns, and they are expected to well-described by a Boltzmann distribution based on a multi-group maximum entropy (MGME) principle \cite{liu2015general,macdonald2018construction,sahai2017adaptive,venturi2020data,sharma2020coarse,Notey2025}. Then, the population fraction of $i$-th rovibrational state within $p$-th group reads:

\begin{equation}
\label{eq:Fp_i}
F_{p}^{i}(T)
= \frac{n_{i}}{n_{p}}
= \frac{g_{i}}{Q_{p}(T)}
  \exp\!\left(-\frac{e_{\text{int}}(i)}{k_{\mathrm{B}} T}\right), \quad i \in I_{p},
\end{equation}

\noindent
where $I_p$ denotes the set of rovibrational states in the $p$-th group. \(n_{p}\) is the total population of group \(p\) (\emph{i.e.}, $n_p=\sum_{i \in I_p} n_i$).
\(Q_{p}(T)\) is the group partition function defined at $T$ and is expressed as:

\begin{equation}
\label{eq:Qp}
Q_{p}(T)
= \sum_{i \in I_{p}} g_{i}
  \exp\!\left(-\frac{e_{\text{int}}(i)}{k_{\mathrm{B}} T}\right),
\end{equation}

\noindent
where \(g_{i}\) is the degeneracy of state \(i\) that includes the nuclear spin multiplicity. For odd $J$, $g_i = 3(2J+1)$, whereas for even $J$, $g_i = (2J+1)$, given that the electronic ground state of $\text{H}_2$ has the degeneracy of one. The group-specific degeneracy, energy, and excitation rate coefficients are correspondingly calculated as:

\begin{equation}
\label{eq:gp}
g_{p} = \sum_{i \in I_{p}} g_{i} ,
\end{equation}

\begin{equation}
\label{eq:Ep}
e_{p} =
\frac{\sum_{i \in I_{p}} g_{i} e_{\text{int}}(i)}
     {g_{p}} ,
\end{equation}

\begin{equation}
\label{eq:kpq}
k_{p \to q}
= \sum_{i \in I_{p}} \sum_{j \in I_{q}}
  k_{i \to j} \, F_{p}^i .
\end{equation}

\noindent
where $q$ also denotes group index. The group-specific dissociation rate coefficient (\emph{i.e.}, $k_{p \rightarrow c}$) can be defined in a similar manner with Eq. (\ref{eq:kpq}) \cite{venturi2020data}.
Then the rovibrational StS master equation in Eqs. (\ref{eq:i_ME}) and (\ref{eq:H_ME}) can be reduced to $p$-th group as:

\begin{equation}
\label{eq:p_ME}
\frac{dn_{p}}{dt}
= \sum_{q} \left( k_{q \to p}\, n_{q}\, n_{\mathrm{H}}- k_{p \to q}\, n_{p}\, n_{\mathrm{H}} \right)+ k_{c \to p}\, n_{\mathrm{H}}^{3}- k_{p \to c}\, n_{q}\, n_{\mathrm{H}} ,
\end{equation}

\begin{equation}
\label{eq:H_ME_p}
\frac{dn_{\mathrm{H}}}{dt}
= -\frac{dn_{\ce{H2}}}{dt}
= 2 \sum_{p} \left( k_{p \to c}\, n_{p}\, n_{\mathrm{H}}- k_{c \to p}\, n_{\mathrm{H}}^{3} \right) .
\end{equation}

\noindent
The obtained group population $n_p$ can be used to reconstruct the rovibrational-specific population $n_i$ using Eq. (\ref{eq:Fp_i}).

In the coarse-graining method, it is critical to define groups that accurately capture the system's physical characteristics. In the previous studies, Energy-Based (EB)~\cite{munafo2015modeling, liu2015general} and Vibration-Specific (VS)~\cite{capitelli2012nonequilibrium, colonna2006reduction} coarse-graining methods have been widely used due to their simplicity. However, the following studies have revealed advanced coarse-graining strategies, such as Graph-Based (GB) \cite{sahai2017adaptive} and Centrifugal-barrier-Based (CB) \cite{venturi2020data} approaches, to more accurately capture energy transfer and dissociation dynamics, respectively.
The GB coarse-graining determines the group definition by evaluating the minimum distances between rovibrational states using excitation/de-excitation rate coefficients. 
The CB coarse-graining leverages the distance from the centrifugal-barrier to define a dissociation-favorable group by taking into account the rotational nonequilibrium. In the present study, a hybrid of those two coarse-graining methods (\emph{i.e.}, GB and CB) is referred to as HCG.

In the recent studies, a concept of HCG has been demonstrated for radical exchange \cite{jo2022rovibrational} and recombination dominant \cite{Notey2025} 0-D isothermal isochoric reactors by applying the GB group for the low-lying energy levels, whereas for the dissociation-dominant high-lying energy states, CB coarse-graining was applied. However, those efforts were limited to a specific heat bath temperature rather than being applicable across a wide range of temperatures. This limitation mainly stems from the way the GB coarse-graining defines the group, in which the graph distance is a function of the rate coefficient, which depends on the heat bath temperature. This results in the absence of a single group definition that can cover a wide range of temperatures.

\begin{figure}[hbt!]
\centering
\includegraphics[width=1\textwidth]{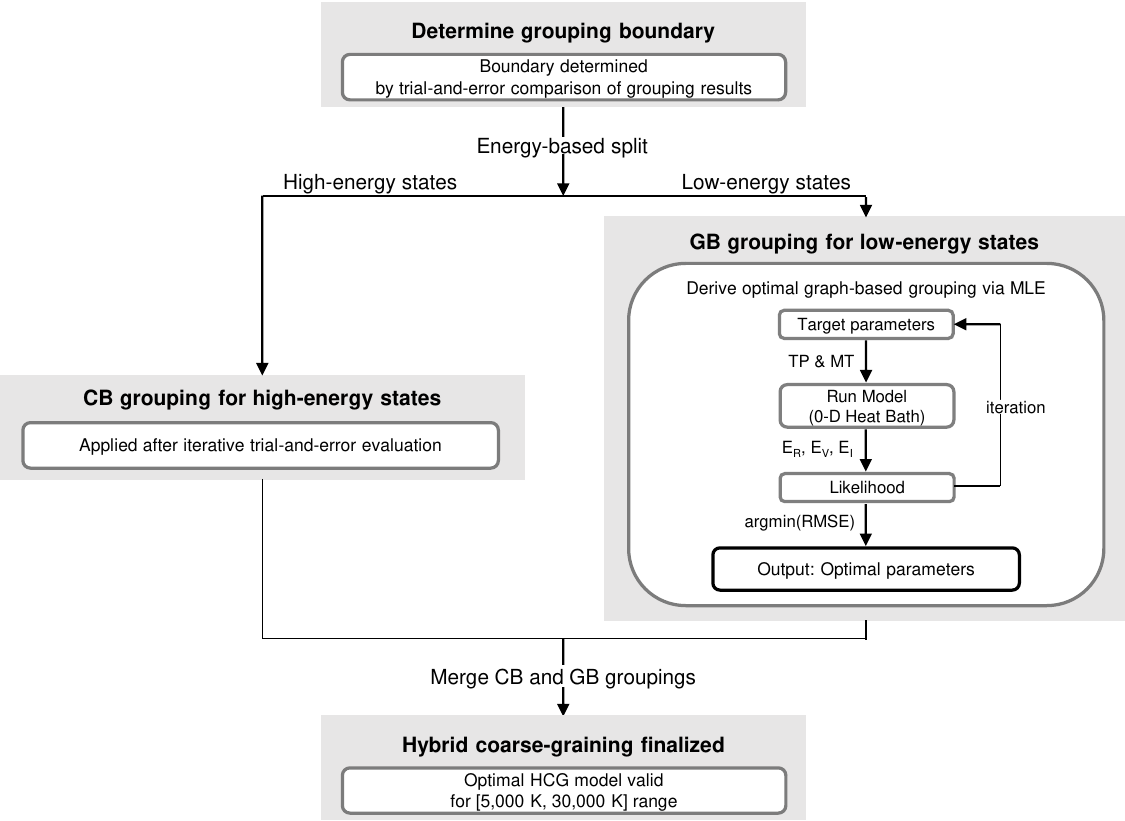}
\caption{Flowchart summarizing the procedure for constructing the HCG model, in which the centrifugal-barrier-based (CB) method is applied to the high-energy states region, while the graph-based (GB) method is employed for the low-energy states region with the optimization via MLE.}
\label{fig:hybrid_group_flowchart}
\end{figure}

To overcome this limitation, in the present study, the HCG approach has been combined with the Maximum Likelihood Estimation (MLE) to define an optimal group definition valid for a wide range of temperatures, 5{,}000~K--30{,}000~K, relevant to hypersonic atmospheric entry flows of the giant planets. Figure~\ref{fig:hybrid_group_flowchart} summarizes the proposed procedure for constructing the present HCG model. The first step is to determine the boundary of the GB and CB coarse-graining. Trial-and-error comparisons on the $\text{H}_2$ mole fraction profiles were used to identify a lower threshold of distance from the centrifugal-barrier over the temperatures that influences the dissociation dynamics. 
Based on this assessment, the CB coarse-graining is applied to the high-lying energy states within 0.062~Eh of the centrifugal-barrier with $e_{\text{int}}(v_{\max},J{=}0)$ zero. Beyond this threshold, the GB coarse-graining approach is applied. The minimum number of CB groups within this distance threshold from the centrifugal-barrier was also determined by trial-and-error comparisons, identifying the minimum number of groups that ensures the $\text{H}_2$ mole fraction profiles are within 0.05 of the absolute error across the entire time domain of the 0-D heat bath simulations compared to the rovibrational StS results, leading to four CB groups determined. 

The second step is to optimize the graph partitioning of the low-lying states over the wide range of temperatures by leveraging the MLE. The graph partitioning was performed using \textsc{infomap}~\cite{mapequation2025software} toolbox by solving the map equation, whereas the MLE was carried out using \textsc{UQpy}~\cite{olivier2020uqpy} library. The map equation is a flow-based graph-partitioning method that exploits the tendency of a random walker to remain within certain regions of a network for extended periods~\cite{rosvall2009map}. In this context, groups are interpreted as modules, and more efficient compression of the walker's trajectory (\emph{i.e.}, transition between rovibrational states) is achieved when the walker tends to stay within modules and transitions between modules are infrequent. This implies that the use of the rovibrational StS rate coefficients (\emph{i.e.}, $k_{i \rightarrow j}$ and $k_{j \rightarrow i}$) in Eq. (\ref{eq:rovib_transfer}) is valid to determine the optimal graph partition \cite{sahai2017adaptive}.

\begin{figure}[hbt!]
\centering
\includegraphics[width=.5\textwidth]{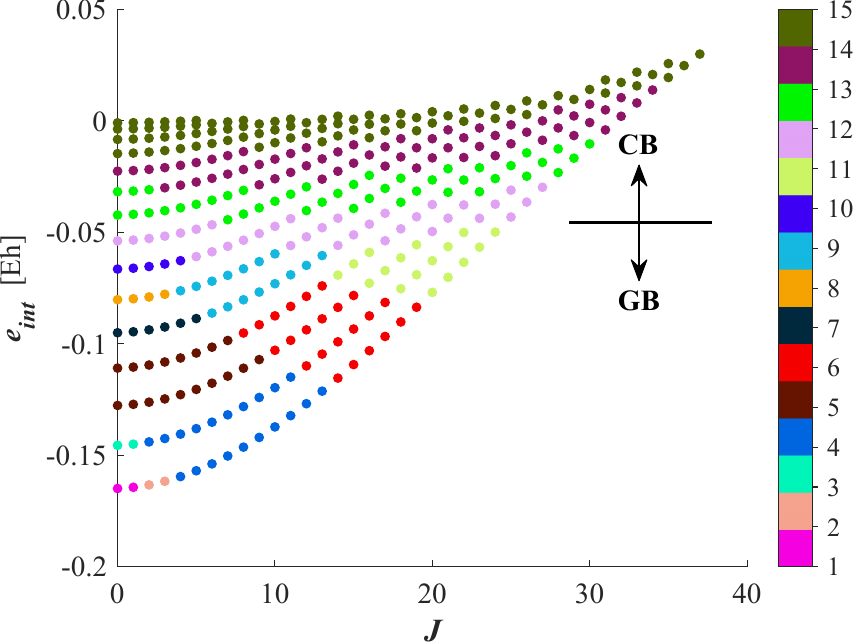}
\caption{Contour of the determined HCG group indices overlayed on the effective diatomic potential of $\text{H}_2\left(\text{X}^1 \Sigma_g^+\right)$. Each dot represents a rovibrational state, and the groups are labeled with assigned colors.}
\label{fig:hybrid_group_contour}
\end{figure}

The MLE was then performed to optimize the teleportation probability (TP) and the Markov time (MT), which are the graph partitioning parameters of the map equation \cite{rosvall2009map}. Both the TP and MT control the degree to which the random walker explores the network in the map equation. The MT represents the time scale over which the random walker explores the network before being encoded; therefore, a smaller MT emphasizes finer graph structures, while a larger value highlights more global structures~\cite{kheirkhahzadeh2016efficient, schaub2012encoding}. 
The MLE is designed to optimize the TP and MT parameters to minimize the root-mean-square-errors (RMSE) of $E_R$, $E_V$, and $E_I$ profiles relative to the rovibrational StS results over time. A Gaussian model is adopted as the likelihood function of the MLE. The probability density functions at individual temporal grid $t$ were treated as independent, and higher weights were assigned to the equilibrium regime of the individual 0-D heat bath simulations to impose physical constraints of the proper thermal relaxation.
The prior ranges for the TP and MT are set to [0, 1] and [0.5, 1.5], respectively, with uniform distributions. The initial estimates for MLE were based on 1{,}000 samples generated using Latin Hypercube Sampling (LHS), and the sample size was established through a convergence study.
As a result, the determined optimized HCG group definition is shown in Fig.~\ref{fig:hybrid_group_contour}. The determined HCG group pattern reflects the state-dependent nature of energy transfer. Around the rovibrational ground state (\emph{i.e.}, $v{=}0$ and $J{=}0$), the transition probability to the nearest neighbor is significantly larger than the other transitions, leading to the finer subdivisions of $p{=}[1,2,3]$ where $p$ denotes the group index. On the other hand, in the high-lying energy region, for example, $p{=}11$, the transition probabilities across the vibrational strands and the rotational states are similar, thus, a larger number of rovibrational states are grouped together. It is worth noting that the total number of groups of the present HCG approach is set to be identical with the number of vibrational states of $\text{H}_2\left(\text{X}^1 \Sigma_g^+\right)$ for fair comparison with the conventional VS coarse-graining model.

\section{\label{sec:4}Results}
The rovibrational-specific master equation analysis is first performed for the 0-D isothermal isochoric heat bath to characterize both the energy transfer and dissociation mechanisms, followed by the comparisons with the reduced-order models. The analysis covers quantities ranging from the microscopic rovibrational state populations to the macroscopic dissociation rate coefficients, relaxation times, and energy loss ratios. Then the model comparison is extended to a 2-D axisymmetric hypersonic flow simulation relevant to Uranus atmospheric entry to address key differences in the flow predictions between the present modified 2T model and the existing 2T approach. It is important to clarify that in the present calculations, $\text{H}_2\left(\text{X}^1\Sigma_g^+\right)$+$\text{H}\left({}^2\text{S}\right)$ collision is the only chemical system considered.

\subsection{\label{sec:results_energy_transfer}Characterization of Energy Transfer Mechanism}

\begin{figure}[hbt!]
\centering
\includegraphics[width=.5\textwidth]{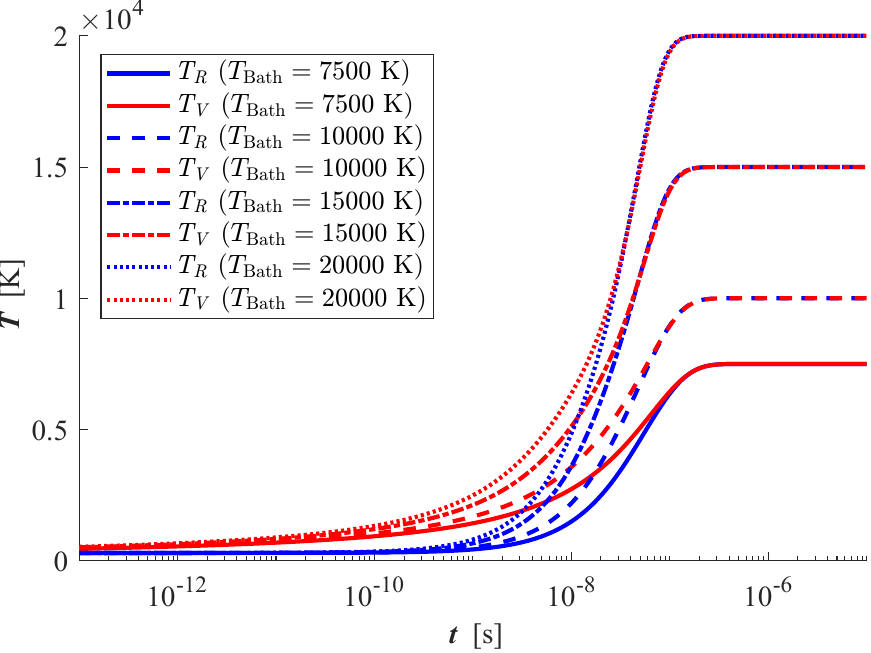}
\caption{Temporal evolutions of the rotational and vibrational temperatures at different heat-bath temperatures (dissociation mechanisms excluded).}
\label{fig:WO_T}
\end{figure}

Figure~\ref{fig:WO_T} shows the time evolution of rotational and vibrational temperatures at heat-bath temperatures ranging from 7{,}500~K to 20{,}000~K, with dissociation reactions excluded.
At all heat-bath temperatures, vibrational relaxation is observed to be faster than rotational relaxation. This behavior becomes more pronounced at higher bath temperatures. Such relaxation dynamics differ from those observed in heavier diatomic species such as \(\ce{N2}\), $\text{O}_2$, and \(\ce{NO}\), where vibrational relaxation is much slower than rotational relaxation~\cite{jo2022rovibrational}. The unique relaxation characteristics of the $\text{H}_2\left(\text{X}^1\Sigma_g^+\right)$+$\text{H}\left({}^2\text{S}\right)$ system were also reported by Kim \textit{et al.}~\cite{kim2009master}. The fast vibrational relaxation of this case implies high transition probabilities in vibrational ladder climbing, arising from the larger rotational energy spacing in $\text{H}_2$, compared to the heavier diatomic species.

\begin{figure}[t!]
    \centering
    \subfigure[Initial target state: $(v,J)=(0,0)$.]
    {
        \includegraphics[width=0.45\textwidth]{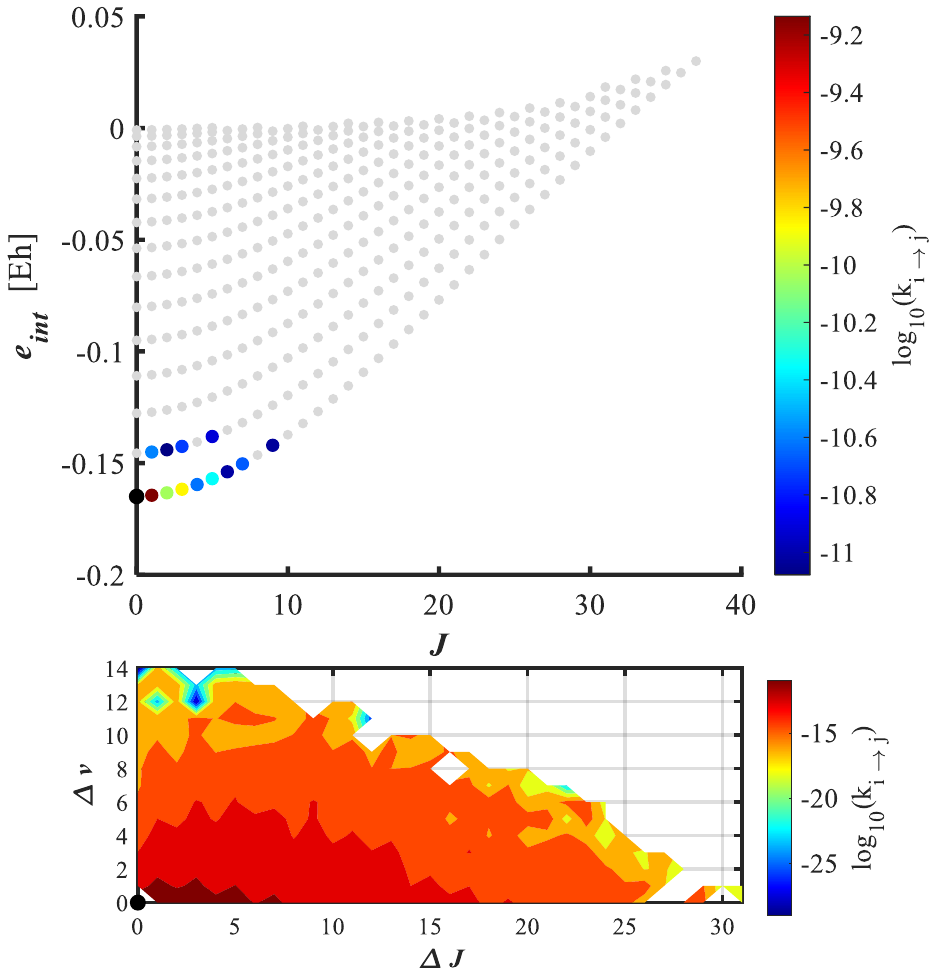}
        \label{inelastic_1}
    }
    \subfigure[Initial target state: $(v,J)=(0,10)$.]
    {
        \includegraphics[width=0.45\textwidth]{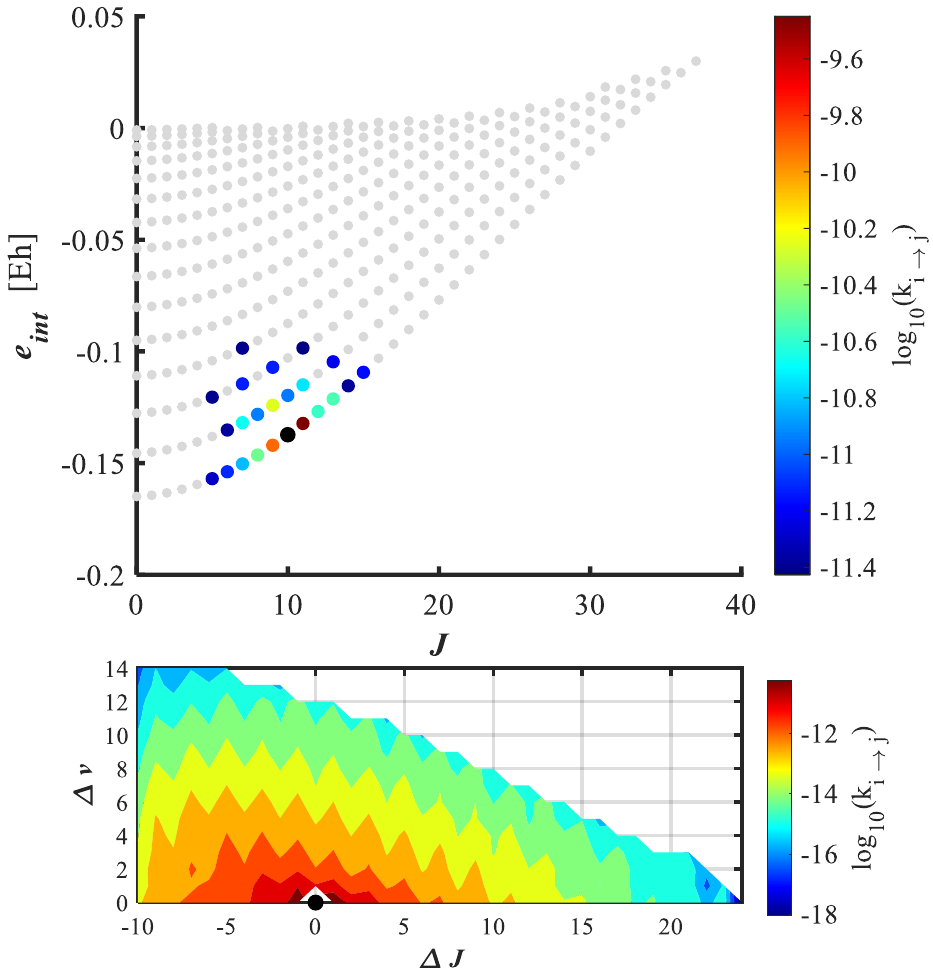}
        \label{inelastic_2}
    }
    \subfigure[Initial target state: $(v,J)=(0,20)$.]
    {
        \includegraphics[width=0.45\textwidth]{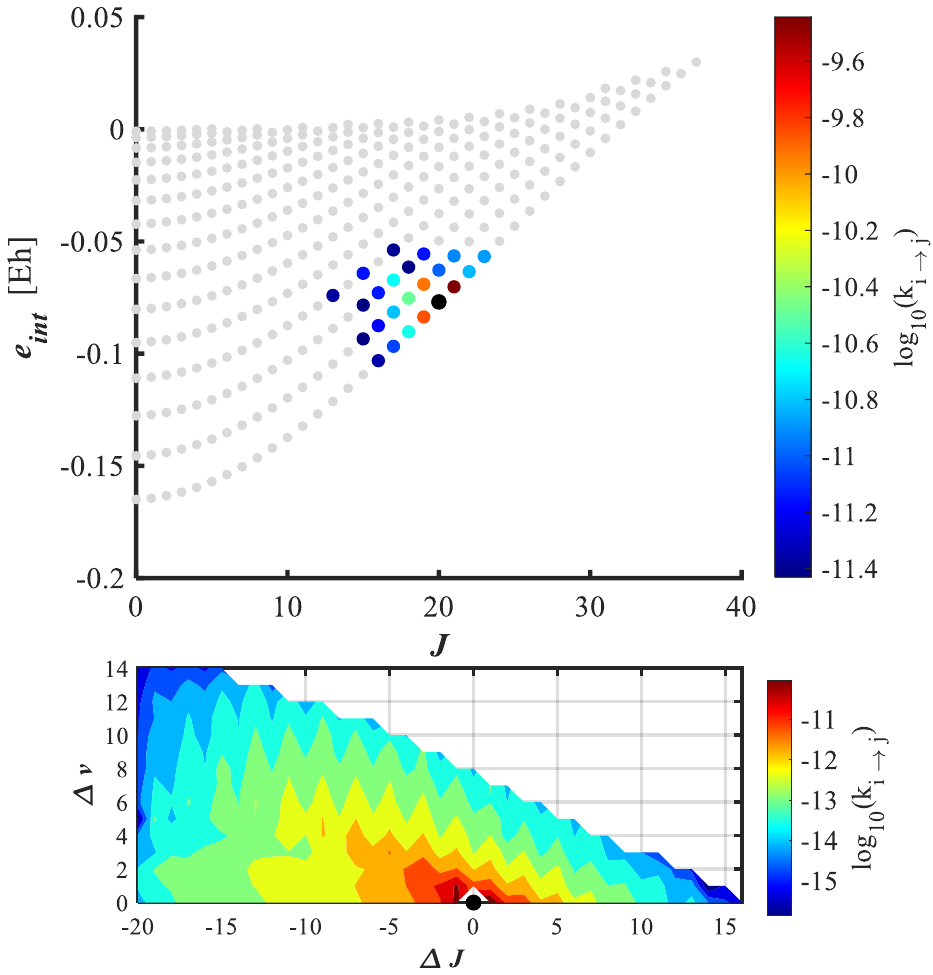}
        \label{inelastic_3}
    }
  \caption{Contour plots of the rovibrational StS inelastic rate coefficients at 10{,}000~K are shown for the initial target state with vibrational quantum number ($v$, $J$), transitioning to ($v'$, $J'$). The top row displays values within two orders of magnitude of the largest rate coefficient on the states. The bottom row presents the rate coefficient magnitude as a function of \(\Delta v\) and \(\Delta J\).}
    \label{inelastic}
\end{figure}

To further support this aspect, Fig.~\ref{inelastic} presents the inelastic rate coefficients for the selected low-lying initial target states, which strongly influence the energy transfer, overlayed on the diatomic potential. The top row shows the magnitudes of the rate coefficients truncated at two orders below the maximum rate coefficient, with the initial target state indicated as black dots. The bottom row shows the corresponding rate coefficient magnitudes as a function of $\Delta J$ and $\Delta v$.
In the present system, the transition probability is significantly large and comparable for both along rotational strand within a vibrational state (\emph{i.e.}, $\left(v, \Delta J\right)$) and across the vibrational quantum number (\emph{i.e.}, $\left(\Delta v, J\right)$) that is further supporting the relaxation trend observed in Fig. \ref{fig:WO_T}.

\begin{figure}[hbt!]
\centering
\includegraphics[width=.5\textwidth]{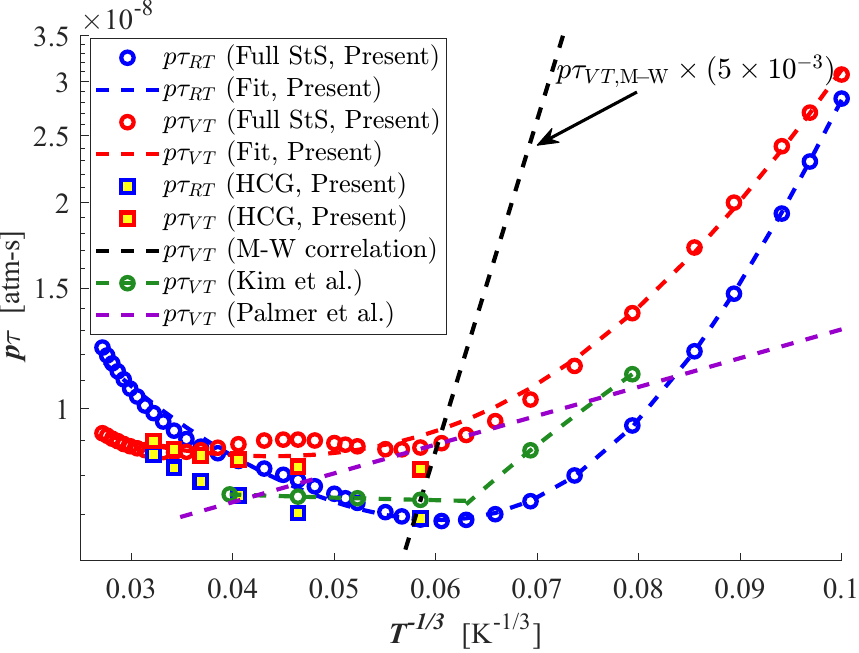}
\caption{Comparisons of the characteristic rotational $(p\tau_{\mathrm{RT}})$ and the vibrational $(p\tau_{\mathrm{VT}})$ relaxation times. The values of $p\tau_{\mathrm{VT}}$ are also compared with existing data~\cite{palmer2014aeroheating,kim2009master,millikan1963systematics},
including those obtained using the Millikan--White correlation, which are shown after being scaled by a factor of $5\times10^{-3}$. The curve-fitted values for the full StS approach and the results obtained from the HCG model are also shown.}
\label{fig:p_tau_comparison}
\end{figure}

Figure~\ref{fig:p_tau_comparison} shows the characteristic rotational \((p\tau_{\mathrm{RT}})\) and the vibrational \((p\tau_{\mathrm{VT}})\) relaxation time obtained from the present master equation study compared to the existing data in literature~\cite{kim2009master, palmer2014aeroheating,millikan1963systematics} as functions of temperature. 
To the authors’ best knowledge, no experimental data exists for the $\text{H}_2\left(\text{X}^1\Sigma_g^+\right)$+$\text{H}\left({}^2\text{S}\right)$ system. The $p\tau_{\text{VT}}$ values obtained using the Millikan--White (M-W) correlation~\cite{millikan1963systematics} differ by several orders of magnitude and are therefore shown in the figure after being scaled by a factor of $5\times10^{-3}$. Taylor \textit{et al.}~\cite{taylor2013estimates} noted that applying the Millikan--White correlation to hydrogen is highly inaccurate. White \textit{et al.}~\cite{white1965vibrational} also reported that the Millikan--White correlation does not accurately predict the vibrational relaxation time for diatomic molecules including hydrogen, attributing this discrepancy to the low molecular weight and the high characteristic vibrational temperature of \(\ce{H2}\). Palmer \textit{et al.}~\cite{palmer2014aeroheating} further stated that the Millikan--White correlation yields excessively large relaxation times for hydrogen by more than several orders of magnitude. Kim \textit{et al.}~\cite{kim2009master} performed StS calculations based on the BKMP2 PES, but the limited temperature range and sparse temperature points led to non-continuous results (\emph{i.e.}, green line with kink), making it challenging to leverage for macroscopic CFD simulations. Palmer \textit{et al.}~\cite{palmer2014aeroheating} refitted the parameters of the Millikan–White formula over the temperature range from 1{,}000 K to approximately 25{,}000 K by replacing the discontinuity in the data of Kim \textit{et al.}~\cite{kim2009master} with a smooth function. Comparing the fitted data of Palmer \textit{et al.}~\cite{palmer2014aeroheating} with the present full StS master equation results highlights the lack of high-temperature correction effects in the existing data, leading to the significant discrepancy in the high-temperature regime.
The results of the present study, obtained from full StS calculations based on the most recent \emph{ab-initio} PES, show longer vibrational relaxation times than those reported in previous studies across most of the temperature range.

To enable the first-order approximation (\emph{e.g.}, the Landau-Teller approximation~\cite{LT_1936}) of the vibrational and rotational energy transfers in macroscopic CFD simulations, we propose new curve-fit parameters for the \(p\tau_{\mathrm{RT}}\) and \(p\tau_{\mathrm{VT}}\) of the rovibrational-specific master equation analysis into the functional forms given in Eqs.~(\ref{eq:tau_total})–(\ref{eq:sigma}). Accordingly, Table~\ref{tab:relaxation-fitting} summarizes the fitting parameters, \(A_{XT}\), \(B_{XT}\), \(C_{XT}\), and \(D_{XT}\), for the relaxation times, where the subscript $X$ denotes either the vibrational ($X{=}V$) and the rotational ($X{=}R$) energy mode. The functional definition of $p\tau_{\text{RT}}$ has the same form as Eqs.~(\ref{eq:tau_total})–(\ref{eq:sigma}).
The importance of the rotational nonequilibrium in predicting hypersonic shock stand-off distances has been discussed by Jo \textit{et al.}~\cite{jo2021prediction}, therefore, we also parameterized the rotational relaxation time along with the one for the vibration. It is important to note that the rotational relaxation of the $\text{H}_2$+H system is comparable with the vibrational contribution, implying its importance in modeling the nonequilibrium chemical-kinetics of the system.
Figure~\ref{fig:p_tau_comparison} also shows the results of the HCG model. Both \(p\tau_{\mathrm{RT}}\) and \(p\tau_{\mathrm{VT}}\) obtained from the HCG model reproduce the full StS results with good agreement and retain the overall behavior. This implies that the present HCG approach can be extended to four-body interactions (\emph{i.e.}, $\text{H}_2\left(\text{X}^1\Sigma_g^+\right)$+$\text{H}_2\left(\text{X}^1\Sigma_g^+\right)$) in future studies to investigate the thermochemical nonequilibrium chemical-kinetics of such systems with tractable computational cost.

\begin{table}[hbt!]
\caption{Summary of the present curve-fitting parameters for \(p\tau_{\mathrm{VT}}\) and \(p\tau_{\mathrm{RT}}\) obtained from the full StS approach over the temperature range of 1{,}000~K--50{,}000~K. The unit of $p\tau$ is [atm$\cdot$s] and the unit of $\sigma$ is [m$^{2}$].}
\label{tab:relaxation-fitting}
\centering
\begin{tabular}{lcc}
\hline\hline
Parameter & \(p\tau_{\mathrm{VT}}\) ($X{=}V$) & \(p\tau_{\mathrm{RT}}\) ($X{=}R$) \\
\hline
\(A_{{XT}}\) & 59.4529 & 78.5243 \\
\(B_{{XT}}\) & 0.0844  & 0.0886 \\
\(C_{{XT}}\) & \(2.618 \times 10^{-22}\) & \(1.549 \times 10^{-21}\) \\
\(D_{{XT}}\) & 0.3981  & 0.2081 \\
\hline\hline
\end{tabular}
\end{table}


\subsection{\label{sec:results_dissociation}Characterization of Dissociation Mechanism}

\begin{figure}[t]
    \centering
    \subfigure[Rotational and vibrational temperatures.]
    {
        \includegraphics[width=0.475\textwidth]{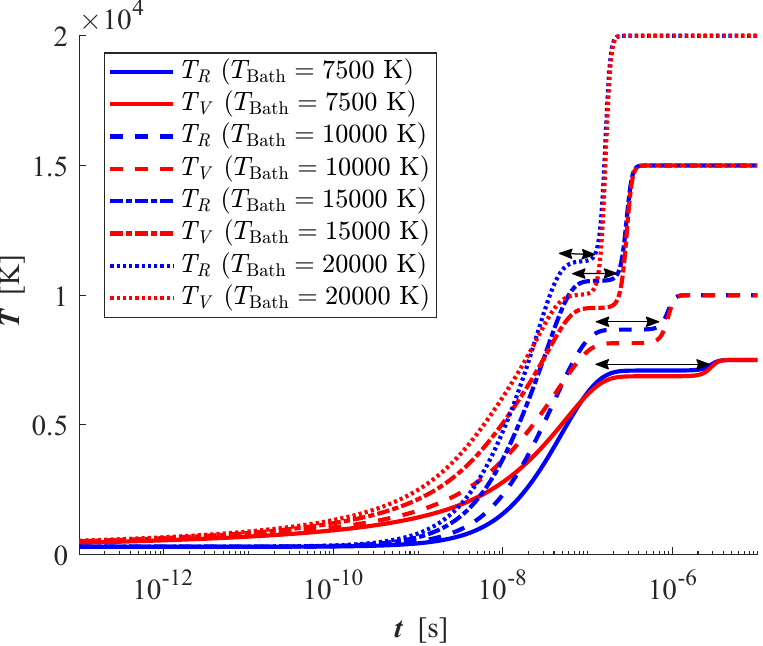}
        \label{fig:W_T}
    }
    \subfigure[Atomic hydrogen mole fractions.]
    {
        \includegraphics[width=0.48\textwidth]{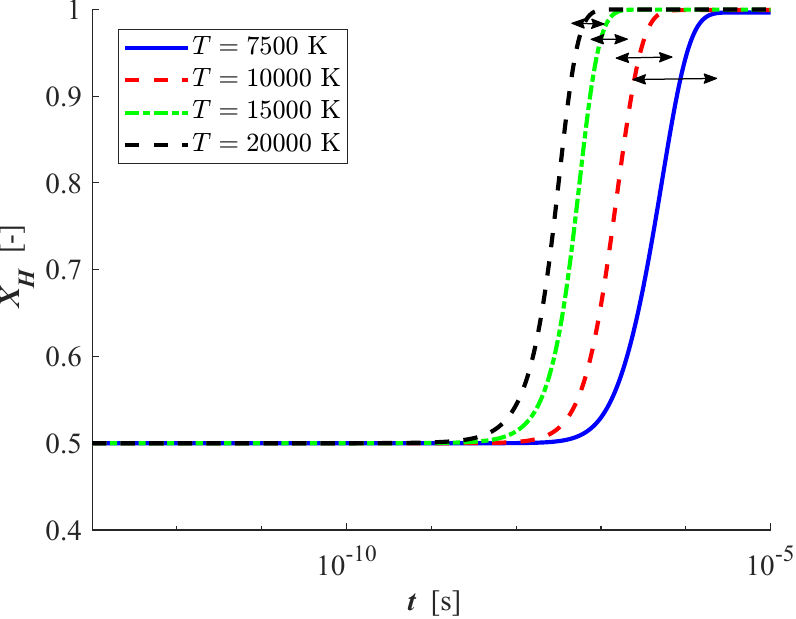}
        \label{fig:X_H}
    }
  \caption{Comparisons of the rotational and vibrational temperatures and the atomic hydrogen mole fraction profiles at different heat-bath temperatures. The bidirectional arrows indicate the QSS periods of $\text{H}_2$.}
    \label{fig:W_T_and_X_H}
\end{figure}

\begin{figure}[t]
    \centering
    \subfigure[Colored by \(v\).]
    {
        \includegraphics[width=0.47\textwidth]{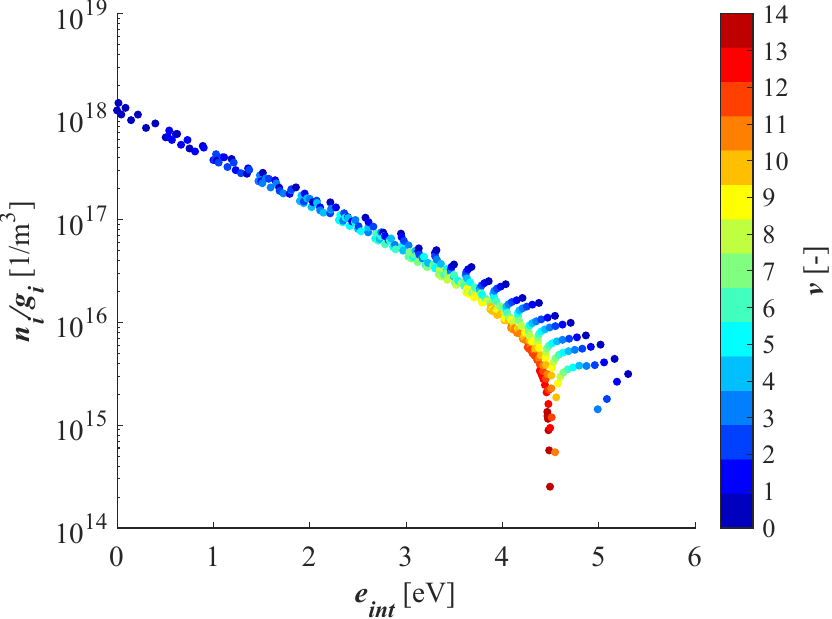}
        \label{fig:QSS_combined_V}
    }
    \subfigure[Colored by $\epsilon$.]
    {
        \includegraphics[width=0.495\textwidth]{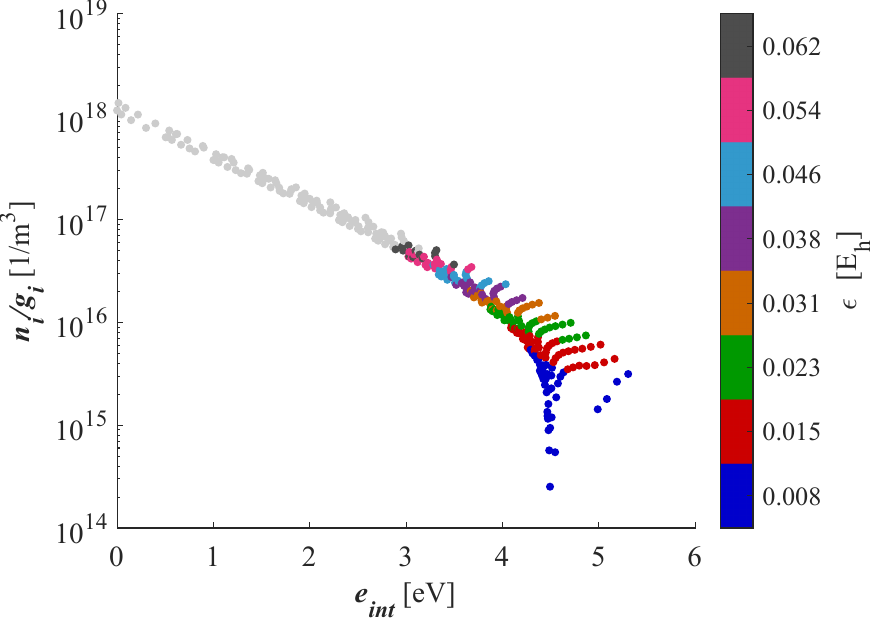}
        \label{fig:QSS_combined_CB}
    }
  \caption{Comparisons of the rovibrational state population distributions at the middle of the QSS region at \(T = 20{,}000\;\mathrm{K}\) along with the different color schemes.}
    \label{fig:QSS_combined}
\end{figure}

Figure~\ref{fig:W_T} shows the time evolution of rotational and vibrational temperatures at heat-bath temperatures ranging from 7,500 to 20,000 K, including the dissociation reactions. The plateau observed in the temperature curves corresponds to QSS periods of $\text{H}_2$, indicated by bidirectional arrows, during which most chemical dissociation occurs. To further understand the behavior of rovibrational states during the period, the QSS population distributions at $T$=20,000 K are shown in Fig.~\ref{fig:QSS_combined}.
In Fig.~\ref{fig:QSS_combined_V}, the states are colored according to the vibrational quantum number \(v\). In Fig.~\ref{fig:QSS_combined_CB}, the same state distribution is colored by the state-specific energy deficit ($\epsilon$) measured from the centrifugal-barrier.
During the QSS period, the states with lower vibrational quantum numbers follow a Boltzmann distribution, whereas the states with higher vibrational quantum numbers (\emph{i.e.}, quasi-bound tails), which significantly contribute to dissociation, do not follow the common Boltzmann trend. These high-\(v\) states follow a common Boltzmann distribution when viewed in terms of the energy deficit as shown in Fig. \ref{fig:QSS_combined_CB}. This indicates that the collision-induced dissociation process in the $\text{H}_2\left(\text{X}^1\Sigma_g^+\right)$+$\text{H}\left({}^2\text{S}\right)$ system cannot be described by a simple vibration-specific behavior, consistent with observations in other chemical systems~\cite{venturi2020data, jo2022rovibrational}. This also demonstrates the validity of applying the centrifugal-barrier-based grouping to the high-energy states in the present HCG approach for the hydrogen systems. Figure~\ref{fig:X_H} presents the time evolution of the mole fraction of atomic hydrogen at the corresponding temperatures of Fig.~\ref{fig:W_T}. It can be seen that most of the dissociation occurs during the QSS periods.
This observation further supports the conventional use of the QSS rate coefficients in CFD analyses~\cite{park1989nonequilibrium} to model chemical production terms due to collisional dissociations.

\begin{figure}[hbt!]
\centering
\includegraphics[width=.5\textwidth]{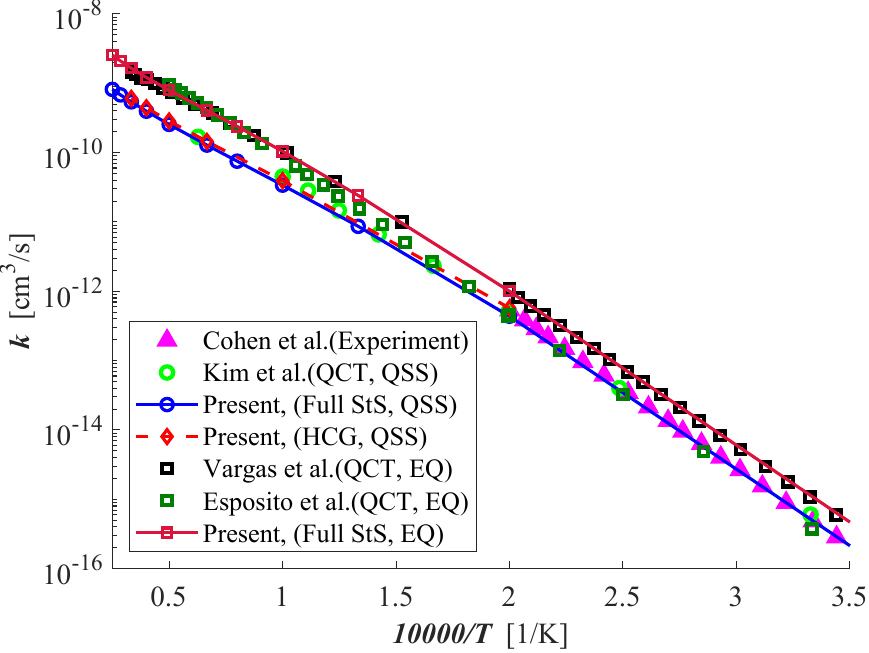}
\caption{Comparison of the present dissociation rate coefficients with the existing data, including Kim \textit{et al.}~\cite{kim2009master}, Vargas \textit{et al.}~\cite{vargas2024state}, Cohen \textit{et al.}~\cite{cohen1983chemical}, and Esposito \textit{et al.}~\cite{ESPOSITO1999636}.}
\label{fig:QSS_themal_rate}
\end{figure}

Figure~\ref{fig:QSS_themal_rate} shows comparisons of the calculated equilibrium (EQ) and the QSS dissociation rate coefficients along with the experimental data. It is generally known that the QSS dissociation rate coefficient is smaller than the thermal rate coefficient~\cite{jo2022rovibrational, panesi2013rovibrational, kim2013state, andrienko2015high, andrienko2016rovibrational}. A similar trend is also observed for the hydrogen system of the present study, as shown in Fig.~\ref{fig:QSS_themal_rate}. This is attributed to the observation in Fig.~\ref{fig:QSS_combined_V}; the high-lying vibrational states are significantly depleted, leading to the lower value of the population-averaged global dissociation rate coefficient during the QSS period (see Eq.~(\ref{eq:k_QSS})). As shown in Fig.~\ref{fig:QSS_themal_rate}, the present QSS rate coefficients are approximately 2.7 times smaller than the present thermal rate coefficients. The present QSS rate coefficient shows good agreement with both the calculations of Kim \textit{et al.}~\cite{kim2009master} and the experimental data of Cohen \emph{et al.}~\cite{cohen1983chemical}, although a slight discrepancy is observed against the rate of Kim \emph{et al.} \cite{kim2009master} around 10,000 K due to the use of different PESs.
We also confirmed that the present HCG approach can offer nearly identical agreement with the full StS result, further demonstrating its applicability to four-body systems. 
The rovibrational StS master equation-derived QSS dissociation rate coefficients by following Eq.~(\ref{eq:k_QSS}) are then curve-fitted to the modified Arrhenius expression given by Eq.~(\ref{eq:k-fitting}). The determined parameters, $A$, $n$, and $C$, are summarized in Table~\ref{tab:k-fitting-params}, which can be adopted by macroscopic CFD simulations.

\begin{table}[t!]
\centering
\caption{Summary of the present curve-fitting parameters for $k_f(T_a)$ obtained from the full StS approach. The unit of $k_f(T_a)$ is [cm$^{3}$/s].}
\label{tab:k-fitting-params}
\begin{tabular}{lc}
\hline\hline
Parameter & Value \\
\hline
\(A\) & \(6.941 \times 10^{-9}\) \\
\(n\) & \(0.0344\) \\
\(C\) & \(50302.81\) \\
\hline\hline
\end{tabular}
\end{table}

\begin{figure}[hbt!]
\centering
\includegraphics[width=.5\textwidth]{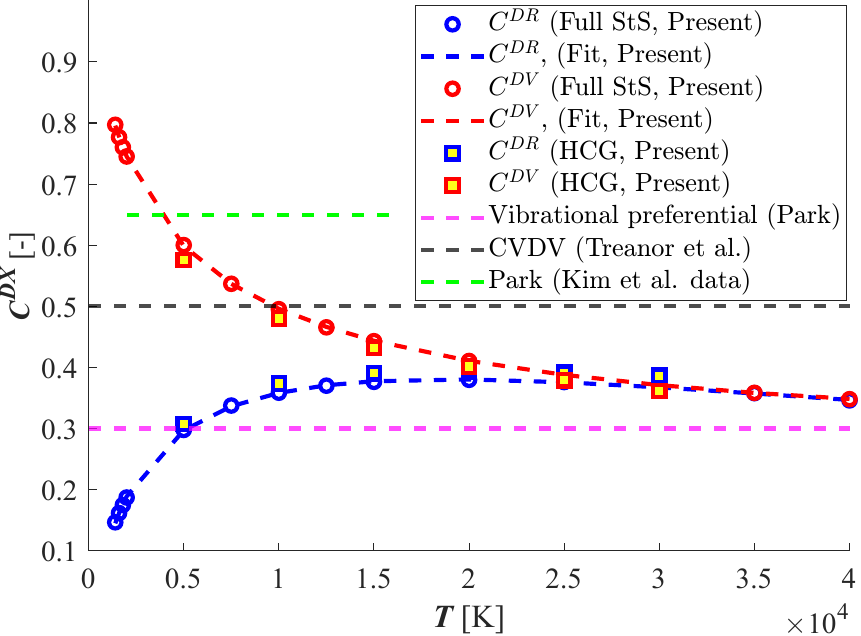}
\caption{Comparisons of the present rotational and vibrational energy loss ratios due to dissociation with the existing data~\cite{park1989nonequilibrium,park2012nonequilibrium,treanor1962effect}.}
\label{fig:energy_loss_ratio}
\end{figure}

Figure~\ref{fig:energy_loss_ratio} compares the present rotational and vibrational energy loss ratios with the literature values, such as the preferential dissociation model proposed by Park~\cite{park1989nonequilibrium}, the coupled vibration–dissociation–vibration (CVDV) model by Treanor and Marrone~\cite{treanor1962effect}, and the representative value of 0.65 suggested by Park~\cite{park2012nonequilibrium} based on Kim \emph{et al.}’s master equation analysis~\cite{kim2009master}.
Compared with the constant values proposed in previous studies, the present full StS results demonstrate that the vibrational energy loss ratio varies significantly with temperature. In addition, the rotational energy loss ratio increases to a non-negligible level as the temperature rises. This behavior indicates that dissociation is dominated by high-$v$ and low-$J$ states at relatively low temperatures, whereas contributions from high-$J$ states in the low-$v$ region become increasingly important at higher temperatures. Moreover, the results of the present HCG approach are shown and agree well with those of the full StS counterpart, further demonstrating its potential for the four-body system. Since the rotational and vibrational energy loss ratios calculated by the present study vary with temperature, curve-fitting is performed to offer a CFD-readable temperature dependent functional format of the obtained master equation analysis using Eq. (\ref{eq:cd-fit}).
The fitting parameters, $K_{1}^X$--$K_{5}^X$ where the superscript $X$ denotes either the rotational ($X{=}R$) or the vibrational ($X{=}V$) contribution, are summarized in Table~\ref{tab:fit_params_loss_ratio}.

\begin{table}[hbt!]
\centering
\caption{Summary of the present curve-fitting parameters of the vibrational and rotational energy loss ratios obtained from the full StS approach. $C^{DV}$ and $C^{DR}$ are dimensionless.}
\label{tab:fit_params_loss_ratio}
\begin{tabular}{lcc}
\hline\hline
Parameter & $C^{DV}$ ($X{=}V$) & $C^{DR}$ ($X{=}R$) \\
\hline
\(K_1^X\) & \(-2.1038 \times 10^{2}\)     & \(-7.9489 \times 10^{2}\)     \\
\(K_2^X\) & \(2.1816\)                    & \(-3.5302\)                  \\
\(K_3^X\) & \(-3.1217 \times 10^{-1}\)    & \(3.0326 \times 10^{-1}\)     \\
\(K_4^X\) & \(1.1282 \times 10^{-6}\)     & \(-2.1939 \times 10^{-5}\)    \\
\(K_5^X\) & \(2.0405 \times 10^{-11}\)    & \(9.6986 \times 10^{-11}\)    \\
Min     & \(0.3481\)                    & \(0.1466\)                    \\
Max     & \(0.7971\)                    & \(0.3798\)                    \\
\hline\hline
\end{tabular}
\end{table}


\subsection{\label{sec:res_models}Comparison of Models}

In this section, results of the various reduced-order models that include the present modified 2T and HCG models, as well as the conventional EB \cite{magin2012coarse} and VS \cite{munafo2012qct} coarse-graining approaches, and the existing 2T model \cite{palmer2014aeroheating,kim2009master,gnoffo1989conservation} in the literature are compared with the reference rovibrational StS method to measure the model performance. The VS model groups states with the same vibrational quantum number, resulting in 15 groups, corresponding to the number of vibrational levels. The EB model defines the group based on uniform internal energy spacing. We equally divided the bound states into 13 groups, with 2 groups assigned to the quasi-bound region, yielding a total of 15 groups, consistent with the VS and HCG models. The existing 2T model implies the following combination: (1) $p\tau_{\text{VT}}$ taken from the work of Palmer \emph{et al.} \cite{palmer2014aeroheating} that does not include the high-temperature correction, (2) QSS dissociation rate coefficient computed by Kim \emph{et al.} \cite{kim2009master}, and (3) a non-preferential dissociation model \cite{gnoffo1989conservation} that is widely used to approximate the chemistry-vibration energy coupling effect in Eq. (\ref{eq:v}). The model comparison ranges from a 0-D isothermal and isochoric reactor to 2-D axisymmetric hypersonic flows relevant to a planetary entry mission to Uranus.

\subsubsection{\label{sec:res_models_0D}0-D Isothermal and Isochoric Reactor}
\begin{figure}[hbt!]
    \centering
    \subfigure[$E_I$.]
    {
        \includegraphics[width=0.45\textwidth]{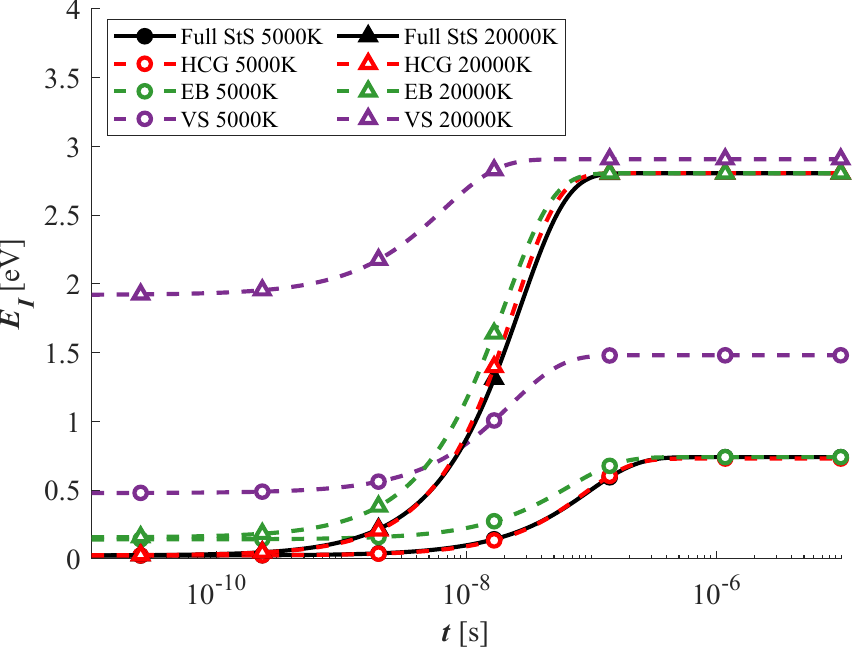}
        \label{fig:models_energy_profiles}
    }
    \subfigure[$T_V$.]
    {
        \includegraphics[width=0.45\textwidth]{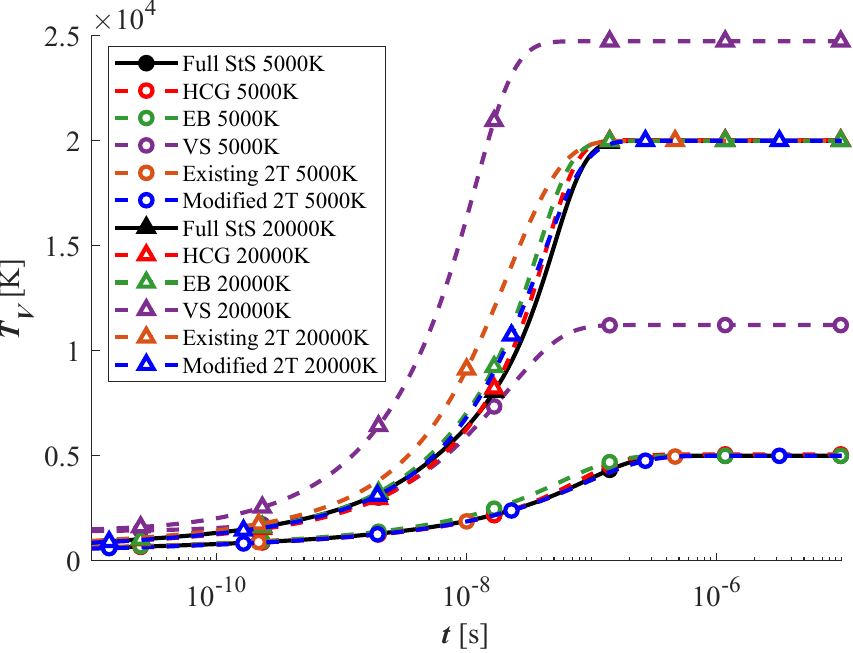}
        \label{T_V_no_diss}
    }
  \caption{Comparisons of the internal energy and vibrational temperature profiles among the HCG, EB, VS, Existing 2T, and Modified 2T models to the full StS results (dissociation mechanisms excluded).}
    \label{energy_T_V_no_diss}
\end{figure}

The condition for the 0-D heat-bath analysis is the same as described in Sec.~\ref{sec:methods_StS}. Figure~\ref{fig:models_energy_profiles} compares the internal energy profiles at the two selected temperatures where the dissociation kinetics are excluded to analyze the energy transfer characteristics of the models. 
Compared with the full StS results, the present HCG model accurately predicts both the nonequilibrium relaxation processes and the post-relaxation equilibrium state over the considered temperature range. This is attributed to the physics-informed coarse-graining of the low-lying energy regime, combined with the MLE-based optimization that enables the application of the model to a wide range of temperatures. 
The conventional EB model overpredicts the internal energy during the initial nonequilibrium evolution, leading to faster relaxation than the full StS method. It is important to note that the VS model fails to reproduce the trend of the full StS result. The degree of failure of the VS model in the present $\text{H}_2\left(\text{X}^1\Sigma_g^+\right)$+$\text{H}\left({}^2\text{S}\right)$ system is significantly more severe than what has been observed for heavier molecular systems, such as $\text{N}_2$+N \cite{munafo2012qct}, $\text{O}_2$+O \cite{venturi2020data}, and air mixtures \cite{su2018state}. This is mainly due to the invalidity of the assumption behind the VS model that rotational levels with a common vibrational quantum number equilibrate rapidly in the hydrogen system, owing to the non-negligible effect of rotational-vibrational-translational (RVT) energy couplings. For the 2T models, the comparison is also made in terms of $T_V$ as shown in Fig. \ref{T_V_no_diss}. The present modified 2T model accurately predicts the vibrational temperature relaxation compared to the full StS counterpart, whereas the existing 2T approach overpredicts the growth rate of $T_V$ at $T$=20,000 K, mainly due to the lack of the high-temperature correction in the collision-limited regime.

\begin{figure}[t!]
    \centering
    \subfigure[HCG (red dots) against the full StS (black dots).]
    {
        \includegraphics[width=0.45\textwidth]{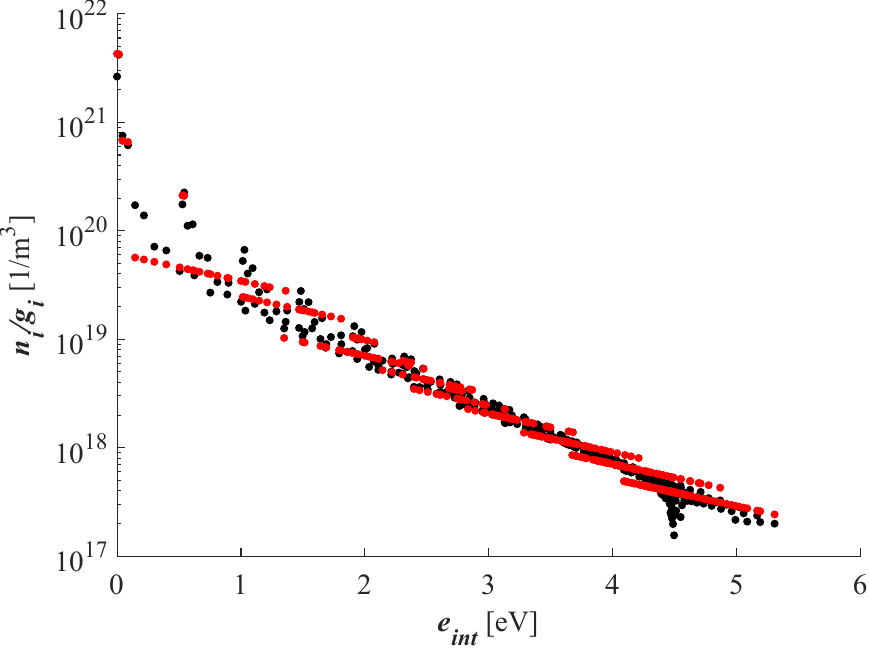}
        \label{fig:recon_pop_woDiss_HCG}
    }
    \subfigure[EB (green dots) coarse-graining against full StS (black dots).]
    {
        \includegraphics[width=0.45\textwidth]{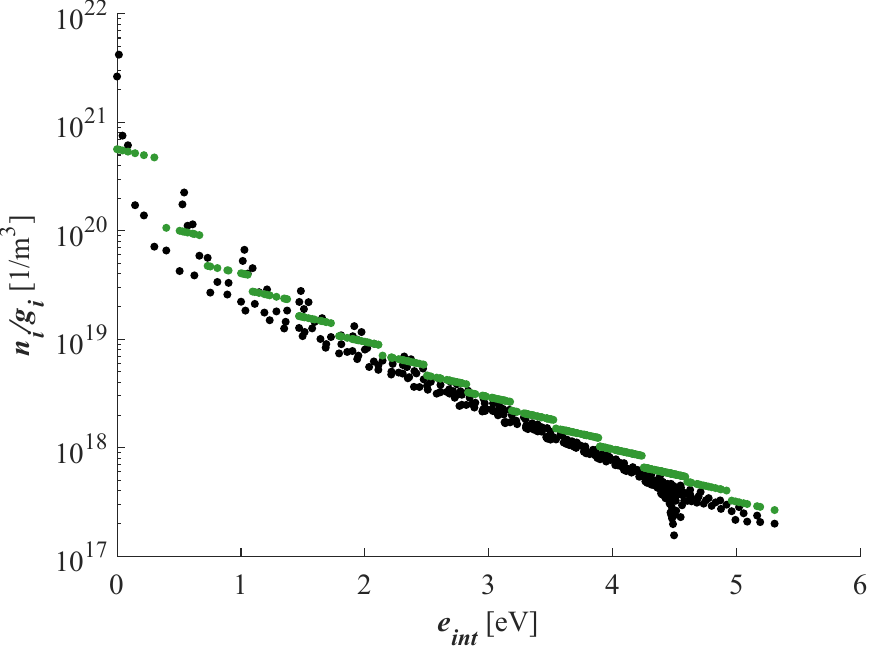}
        \label{fig:recon_pop_woDiss_EB}
    }
    \subfigure[VS (purple dots) coarse-graining against full StS (black dots).]
    {
        \includegraphics[width=0.45\textwidth]{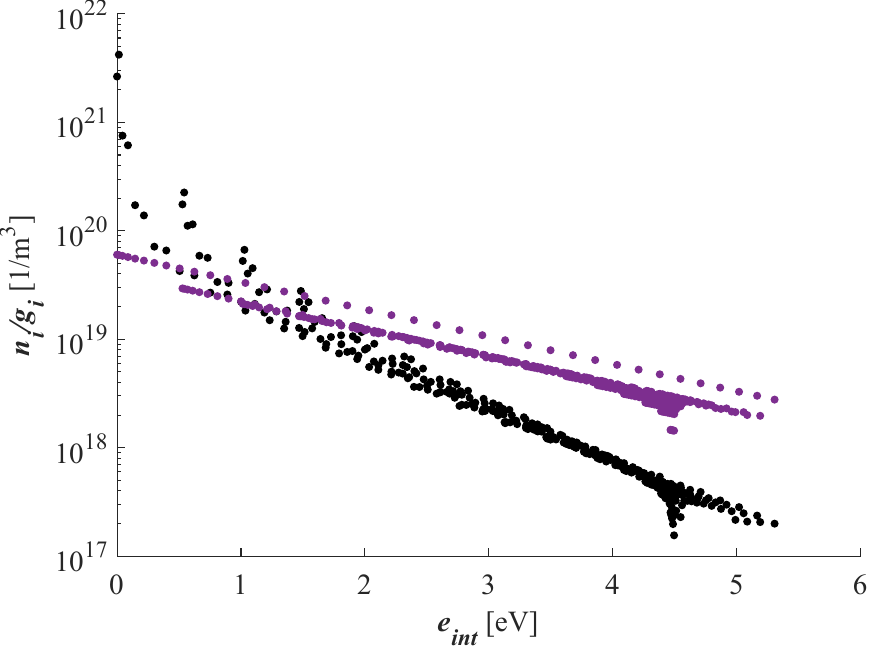}
        \label{fig:recon_pop_woDiss_VS}
    }
  \caption{Comparisons between the full StS results and the reconstructed population distributions of the three coarse-graining strategies at $t{=}10^{-8}\,\mathrm{s}$ of the heat-bath temperature of 20{,}000~K (dissociation mechanisms excluded).}
    \label{fig:recon_pop_woDiss}
\end{figure}

By comparing rovibrational state population distributions at a given time instant in the nonequilibrium region, the model accuracy can be further assessed. Figure \ref{fig:recon_pop_woDiss} presents the instantaneous rovibrational state population distributions of the three different coarse-graining models against the full StS approach at $t{=}10^{-8}$ s of the heat-bath temperature of 20{,}000~K, where the dissociation is excluded. For the coarse-graining models, the population distributions are reconstructed to the rovibrational-specific resolution by imposing the Maxwell-Boltzmann distribution at $T$ within each group using Eq. (\ref{eq:Fp_i}). During the nonequilibrium energy transfer, the fork-like structure of the low-lying energy states is distinct and governs the evolution of the internal energy. The heads of the forks are accurately captured by the HCG model as shown in Fig. \ref{fig:recon_pop_woDiss_HCG}. The EB coarse-graining model reasonably reproduces the low-lying energy states, however, it fails to capture the heads of the forks, leading to the overprediction of the internal energy relaxation, as shown in Fig. \ref{fig:models_energy_profiles}. The high-lying energy state population distribution is accurately reconstructed by the HCG model, whereas the EB method exhibits a slight overprediction, leading to faster excitation of these levels. The reconstructed distribution of the VS coarse-graining model significantly deviates from the full StS counterpart, especially for the fork-like structure near the ground level and for the high-lying energy state populations that are considerably overpredicted, resulting in the excessively faster relaxation as shown in Fig. \ref{fig:models_energy_profiles}.

\begin{figure}[hbt!]
    \centering
    \subfigure[$E_I$.]
    {
        \includegraphics[width=0.45\textwidth]{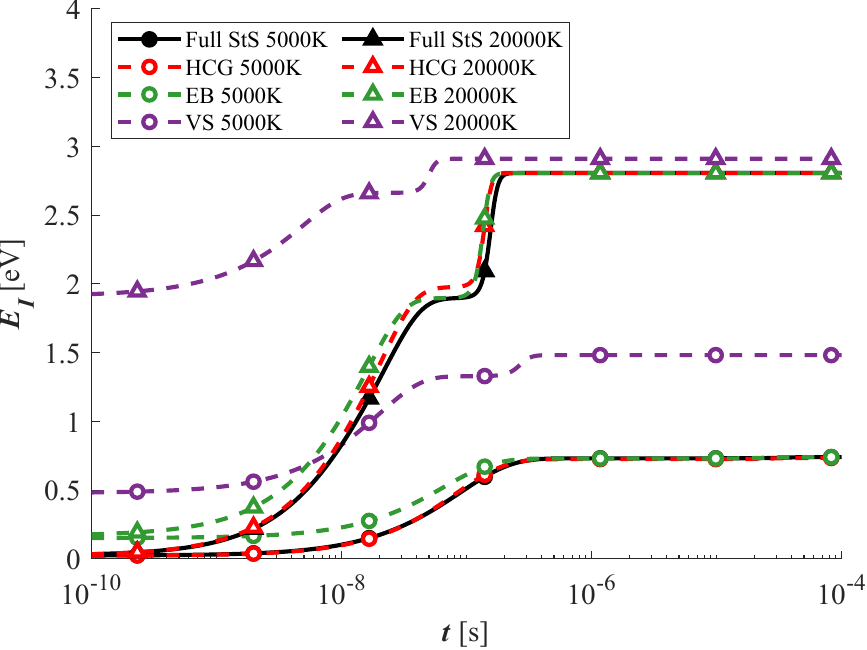}
        \label{fig:models_H2_combined_ei}
    }
    \subfigure[\ce{H2} mole fractions.]
    {
        \includegraphics[width=0.45\textwidth]{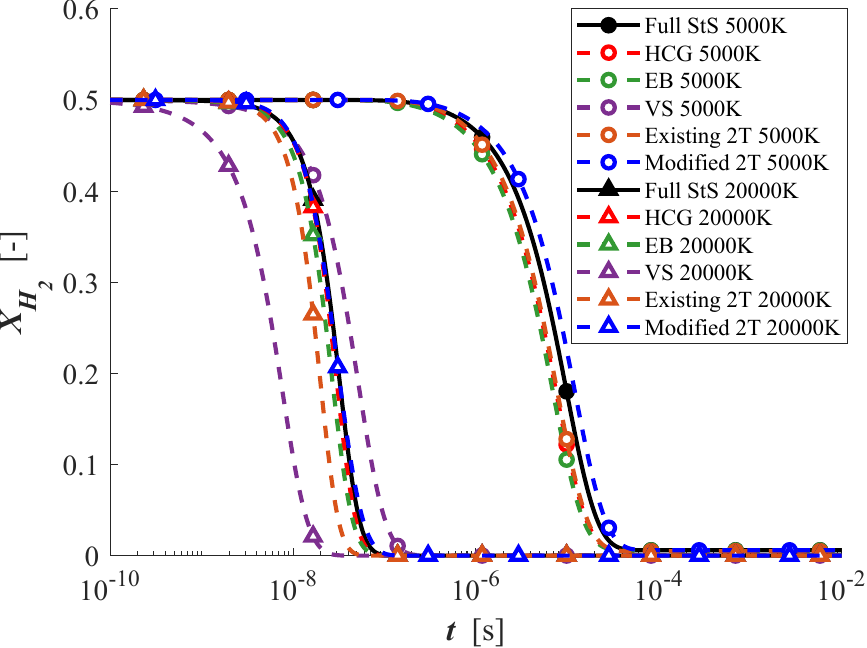}
        \label{fig:models_H2_combined_mole_frac}
    }
    \subfigure[Absolute error of \ce{H2} mole fractions.]
    {
        \includegraphics[width=0.45\textwidth]{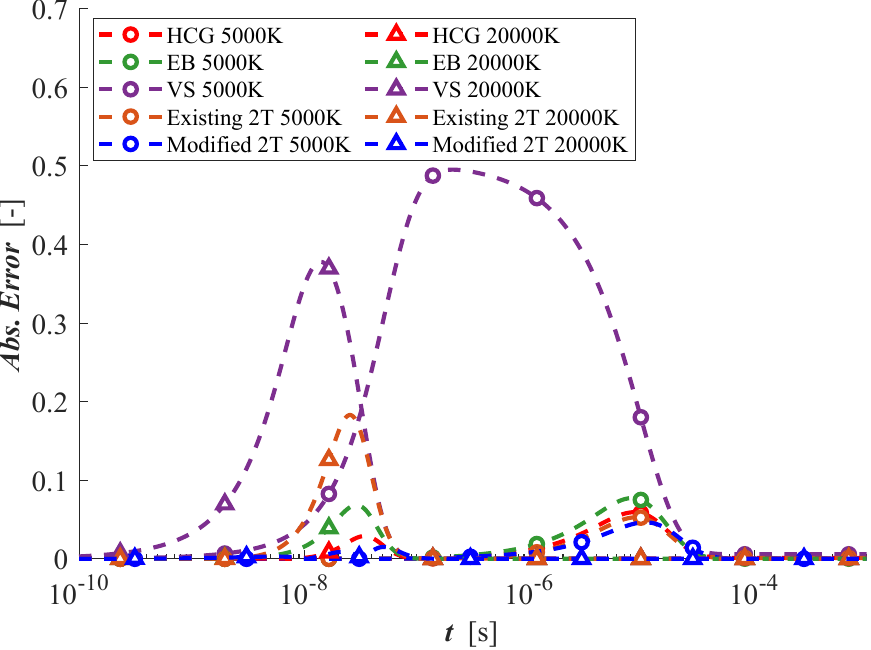}
        \label{fig:models_H2_combined_error}
    }
    \subfigure[$T_V$.]
    {
        \includegraphics[width=0.45\textwidth]{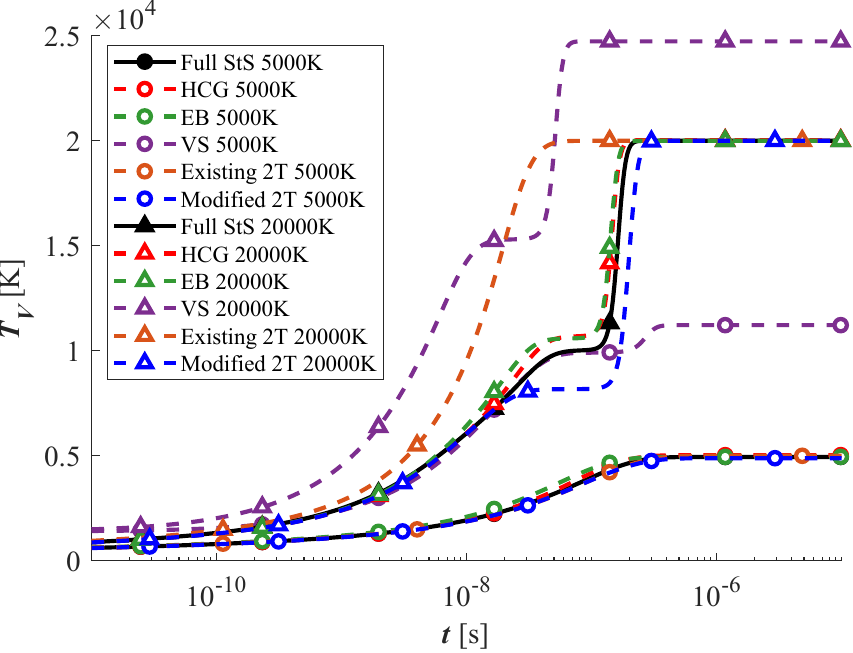}
        \label{fig:models_H2_combined_T_V}
    }
  \caption{Comparisons of the profiles of the internal energy, \ce{H2} mole fractions with absolute errors, and vibrational temperature among the HCG, EB, VS, Existing 2T, and Modified 2T models to the full StS results (dissociation mechanisms included).}
    \label{fig:models_H2_combined}
\end{figure}

Figure~\ref{fig:models_H2_combined} compares the internal energy, the H$_2$ mole fraction, and the vibrational temperature profiles of each model for the dissociation reaction enabled. The VS coarse-graining model predicts significantly faster onset time of the dissociation, as shown in Fig. \ref{fig:models_H2_combined_mole_frac}, compared to the other models. Also, the QSS plateau of the internal energy (\emph{i.e.}, the flat regions of Fig. \ref{fig:models_H2_combined_ei} prior to the equilibrium) of the VS model is formed at the higher energy because of the excessively active internal energy excitation in the initial phase, in which the bound-bound transition governs the chemical-kinetic process, leading to the faster dissociation onset at the higher internal energy state. This observation can be qualitatively justified by the results presented in Fig. \ref{fig:models_energy_profiles} and will be further analyzed in the following paragraphs.
The EB coarse-graining model provides reasonable agreement with the full StS results, except for the overprediction of the internal energy during the initial phase. In particular, it reasonably captures the dissociation dynamics. The HCG model proposed by the present study reproduces the full StS results with the smallest errors for both the internal energy and the $\text{H}_2$ mole fraction profiles, with the errors decreasing further as the heat-bath temperature increases, as shown in Fig. \ref{fig:models_H2_combined_error}. 

The present modified 2T model accurately predicts the $\text{H}_2$ mole fraction and the $T_V$ profiles compared to the full StS counterparts, except for a slight discrepancy of the $T_V$ during the QSS period of 20,000 K heat bath as shown in Fig. \ref{fig:models_H2_combined_T_V}. In addition, its level of error in the prediction of $X_{\text{H}_2}$ is as low as the HCG approach, as shown in Fig. \ref{fig:models_H2_combined_error}. This comparison demonstrates that the proposed modified 2T model accurately predicts both the energy transfer and the dissociation dynamics of the $\text{H}_2$+H system, compared with the full StS and the proposed HCG methods.
On the other hand, the existing 2T model predicts faster dissociation than the present modified 2T approach, due to the combined effect of overestimating the energy transfer (see Fig. \ref{T_V_no_diss}) and the failure to establish a QSS period, as shown in Fig. \ref{fig:models_H2_combined_T_V} for $T$=20,000 K heat bath. In particular, the failure of establishing the QSS period stems from the wrong energy balance on the right-hand side of Eq. (\ref{eq:v}), especially for $\Omega_{c-V}$ by using the non-preferential dissociation model.

\begin{figure}[b!]
    \centering
    \subfigure[HCG (red dots) against the full StS (black dots).]
    {
        \includegraphics[width=0.45\textwidth]{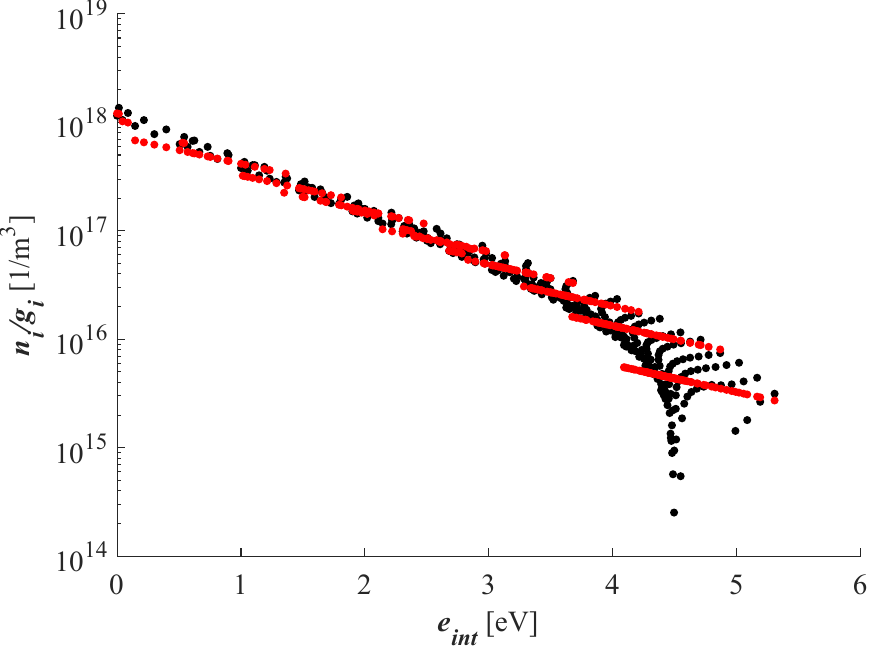}
        \label{fig:recon_pop_all_HCG}
    }
    \subfigure[EB (green dots) coarse-graining against full StS (black dots).]
    {
        \includegraphics[width=0.45\textwidth]{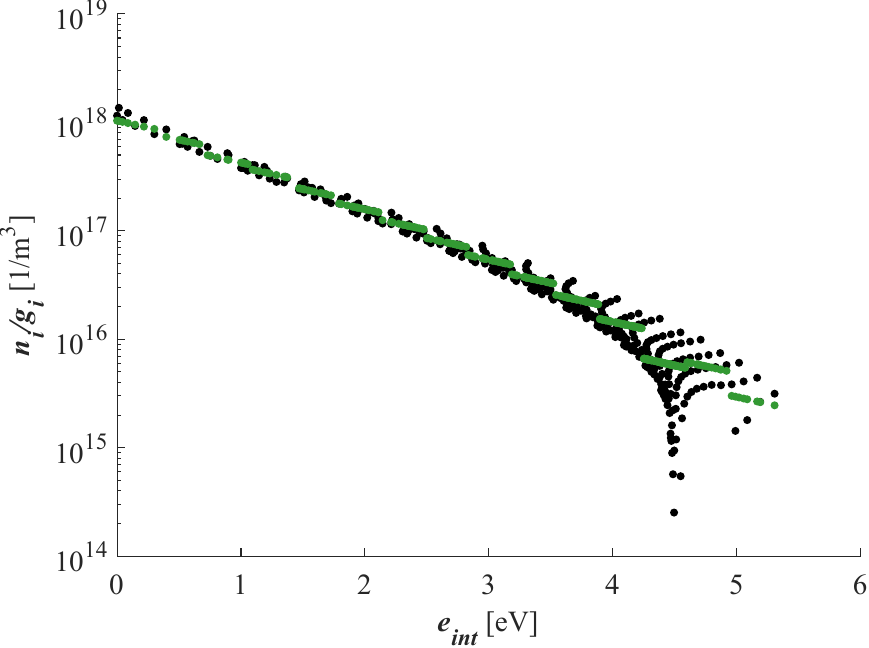}
        \label{fig:recon_pop_all_EB}
    }
    \subfigure[VS (purple dots) coarse-graining against full StS (black dots).]
    {
        \includegraphics[width=0.45\textwidth]{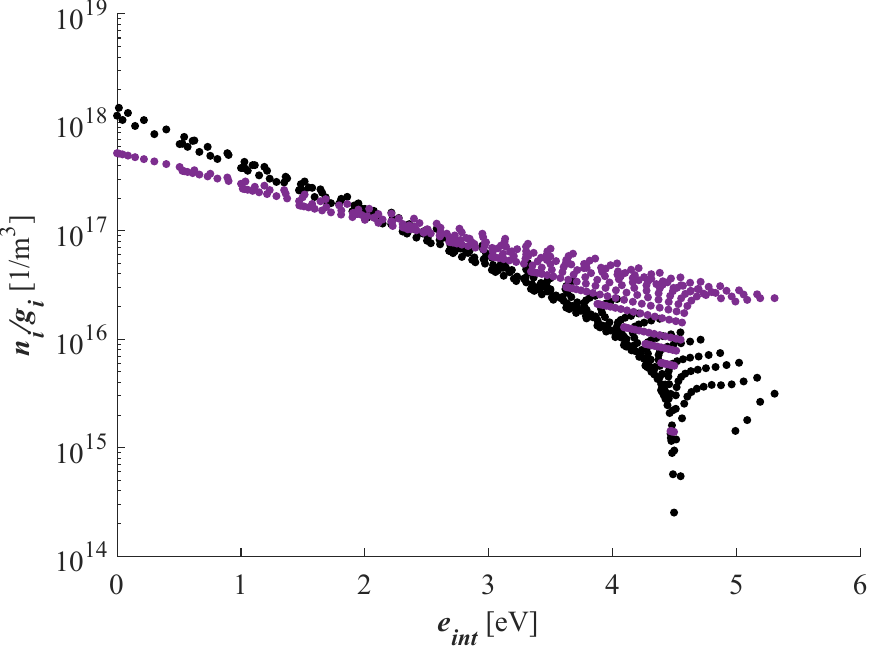}
        \label{fig:recon_pop_all_VS}
    }
  \caption{Comparisons between the full StS results and the reconstructed population distributions of the three coarse-grouping strategies in the QSS region at the heat-bath temperature of 20{,}000~K.}
    \label{fig:recon_pop_all}
\end{figure}

To develop an advanced reduced-order model valid for the dissociation dynamics, it is important to capture the rovibrational state population distribution during the QSS period in which the majority of chemical dissociation occurs \cite{jo2022rovibrational}. 
Figure~\ref{fig:recon_pop_all} shows the reconstructed populations of each coarse-graining method in the QSS region along with the full StS counterpart. It is important to note that the populations are extracted at different times to ensure that the same macroscopic $\text{H}_2$ mole fraction ($X_{\text{H}_2}$=0.003) is obtained across the models, while all population distributions belong to their own QSS periods. Both HCG and EB models offer reasonably accurate reconstructed QSS population distributions compared with the full StS result, except for the highest vibrational quantum states, which are most depleted near the dissociation limit of 4.477 eV. The HCG model reproduces the depletion slightly better than the EB coarse-graining near the tail. 
As expected, the VS model fails to reproduce the depletion of high-lying states, leading to the overprediction of the population distribution, especially for the high-$J$ and low-$v$ states. In addition, the rovibrational ground state population is underestimated by the VS approach due to the overprediction of the thermal relaxation (see Fig. \ref{energy_T_V_no_diss}). This confirms that the VS approach fails to reproduce both the bound-bound energy transfer and the dissociation characteristics.

Inspired by the complete failure of the VS coarse-graining approach, we carried out additional analysis to better understand the root of the issue. 
For \ce{H2}, the importance of rotational nonequilibrium has been discussed in several previous studies. For heavier molecules such as \ce{N2}, \ce{NO}, and \ce{O2}, rotational energy relaxation is much faster than the vibrational relaxation, and RVT energy transfers thus have often been neglected without significant loss of accuracy. In contrast, \ce{H2} has rotational energy gaps that are comparable to those of the vibrational states~\cite{mandy1999state,kim2009master,kim2010master,kim2015rovibrational,furudate2006coupled}. As a result, strong coupling can occur among the rotational, vibrational, and translational energy modes and nonequilibrium effects of both rotation and vibration are important in \ce{H2}~\cite{sharma1991rate,sharma1994rotational,kim2009master,kim2010master,kim2015rovibrational}.
Through Monte Carlo trajectory simulation at 4,500 K, Haug \textit{et al.}~\cite{haug1987monte} argued the importance of the rotational nonequilibrium for each vibrational quantum state of $\text{H}_2$ for developing an advanced reduced-order model for the dissociation process.
In \ce{H2} recombination, the final rotational state can influence the reaction rate by two to three orders of magnitude~\cite{esposito2009selective,capitelli2011molecular,capitelli2016reactivity}. In addition, rotational states in \ce{H2} modify the effective molecular potential through the centrifugal term, indicating strong coupling between vibration and rotation~\cite{colonna2012statistical}.
To the authors' best knowledge, however, no study has demonstrated the failure of the VS coarse-graining strategy for the hydrogen system through detailed master equation analyses. Figures \ref{energy_T_V_no_diss}--\ref{fig:recon_pop_all} of the present study quantitatively demonstrate the failure of the VS coarse-graining method for the $\text{H}_2\left(\text{X}^1\Sigma_g^+\right)$+$\text{H}\left({}^2\text{S}\right)$ system, for the first time, further supporting the previous findings.

\begin{figure}[ht!]
    \centering
    \subfigure[Combined GB and VS coarse-graining.]
    {
        \includegraphics[width=0.45\textwidth]{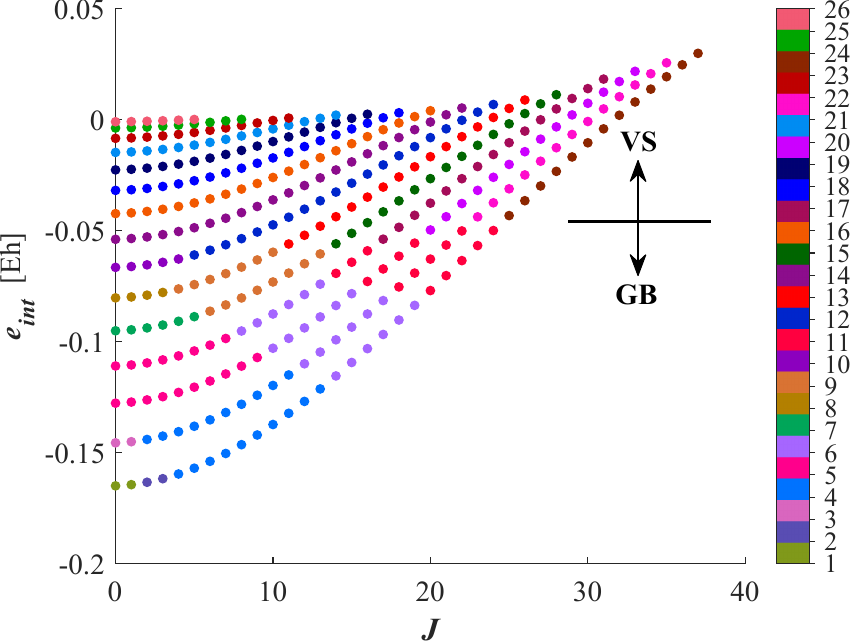}
        \label{VS_mix_contour_GB-VS}
    }
    \subfigure[Combined VS and CB coarse-graining.]
    {
        \includegraphics[width=0.45\textwidth]{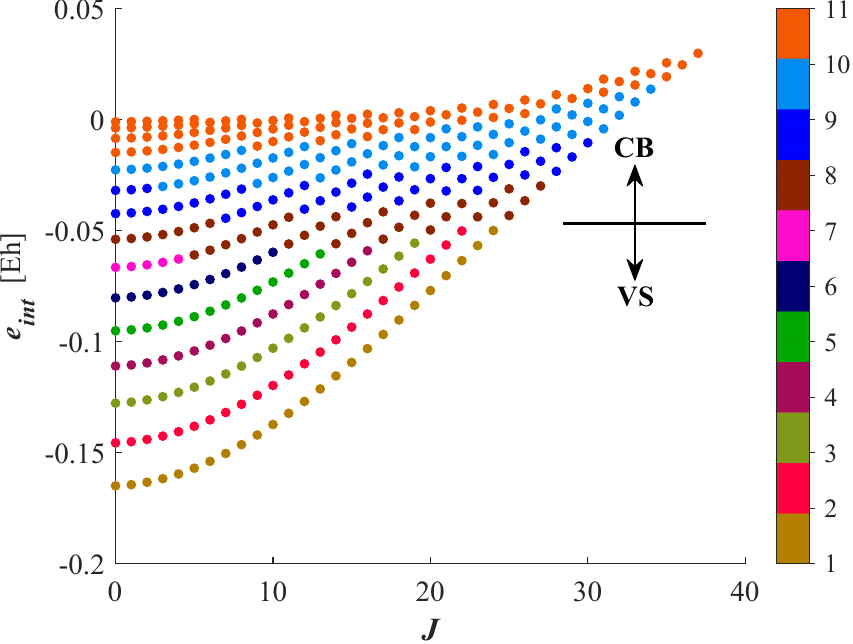}
        \label{VS_mix_contour_VS-CB}
    }
  \caption{Definitions of the additional coarse-graining methods in which the vibration-specific method is applied to the high-lying (figure (a)) and to the low-lying (figure (b)) energy regions of the original HCG approach.}
    \label{VS_mix_contour}
\end{figure}

\begin{figure}[h!]
    \centering
    \subfigure[$E_I$ (dissociation mechanisms excluded).]
    {
        \includegraphics[width=0.45\textwidth]{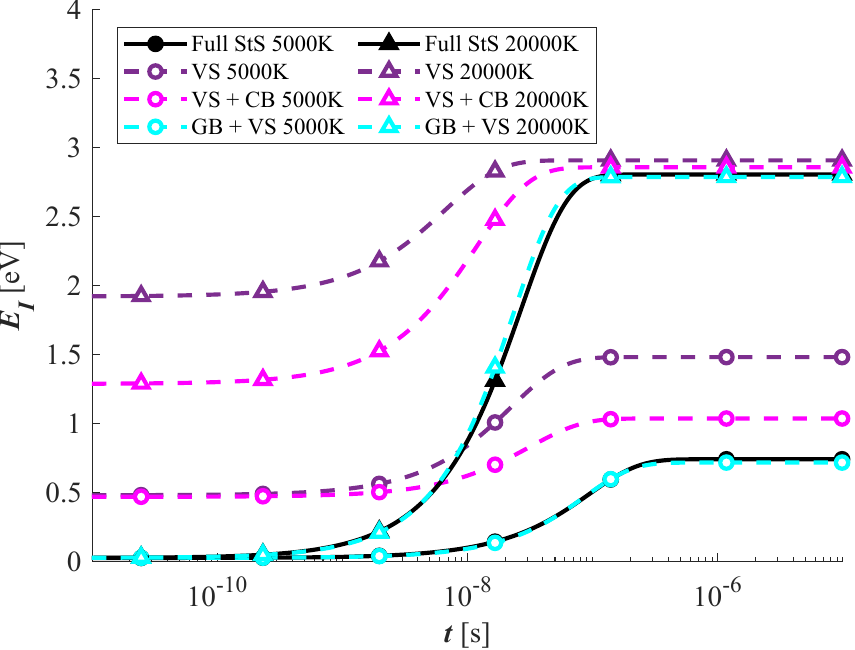}
        \label{fig:VS_mix_energy_H2_mole_fraction_ei_wo_diss}
    }
    \subfigure[$E_I$ (dissociation mechanisms included).]
    {
        \includegraphics[width=0.45\textwidth]{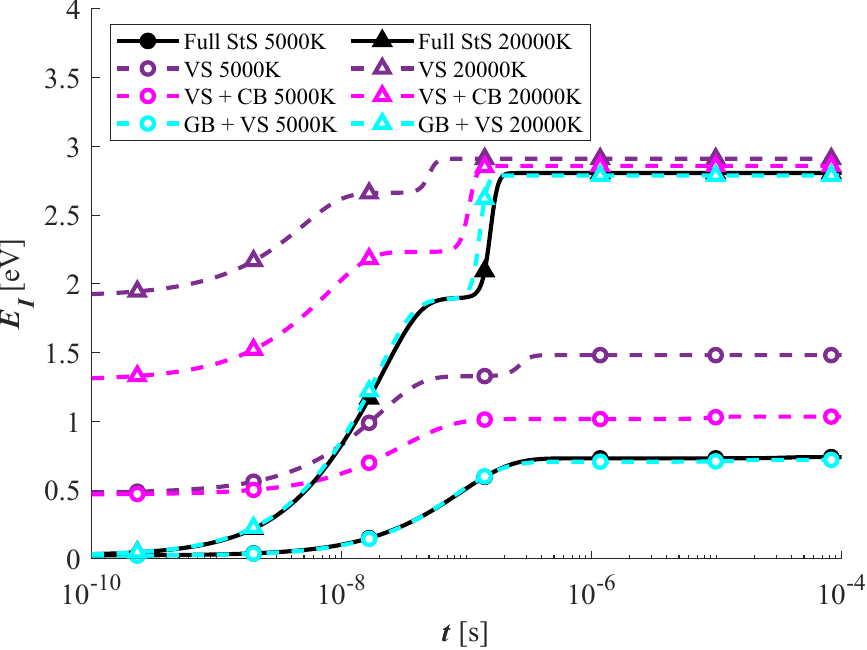}
        \label{fig:VS_mix_energy_H2_mole_fraction_ei_w_diss}
    }
    \subfigure[\ce{H2} mole fractions.]
    {
        \includegraphics[width=0.45\textwidth]{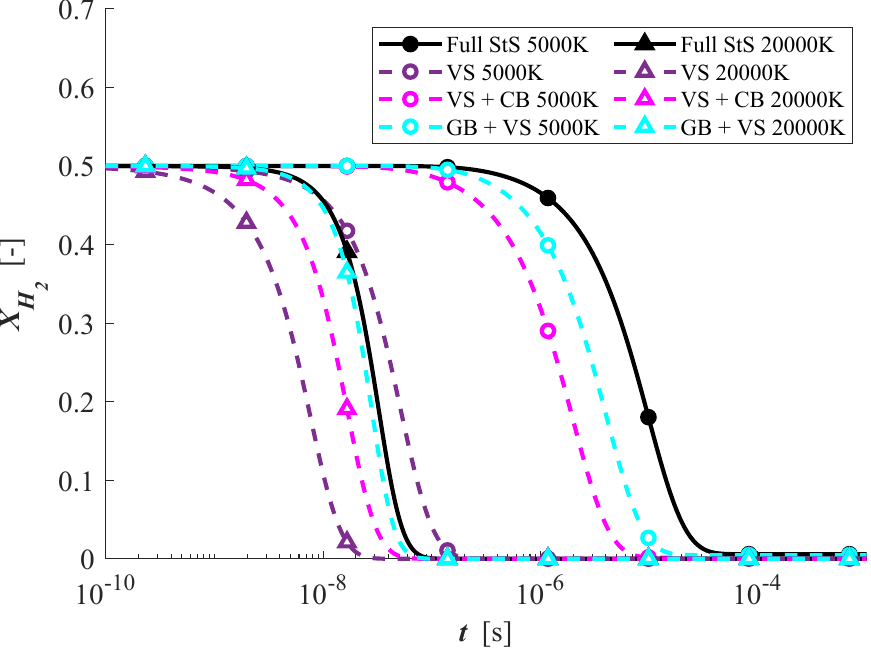}
        \label{fig:VS_mix_energy_H2_mole_fraction_H2_molefrac}
    }
  \caption{Comparisons of the internal energy and \ce{H2} mole fraction profiles among the VS, VS+CB, and GB+VS models against the full StS.}
    \label{fig:VS_mix_energy_H2_mole_fraction}
\end{figure}

In addition, the present study further analyzes the root of the VS coarse-graining model's failure by separating the coarse-graining domains into energy-transfer and dissociation-dominant regions to better pinpoint the source of the inability.
Figure \ref{VS_mix_contour} shows the definition of two extra coarse-graining models for this purpose. Figure \ref{VS_mix_contour_GB-VS} represents a combined graph-based (GB) and VS coarse-grainings, respectively, for the low-lying and the high-lying energy domains. The GB model used here is identical to that used for the HCG model and was determined by the MLE-based optimization. The high-lying dissociation-dominant regime is grouped based on $v$. This first extra model is noted as GB+VS hereafter. Figure \ref{VS_mix_contour_VS-CB} shows the second extra coarse-graining method, which is a hybrid of the VS and the centrifugal-barrier-based (CB) coarse-grainings, respectively, for the low-lying and the high-lying energy states. This is denoted as VS+CB hereafter. These two extra coarse-graining models are expected to provide further intuition on the failure of the VS coarse-graining approach.

Figure~\ref{fig:VS_mix_energy_H2_mole_fraction} compares the internal energy and $\text{H}_2$ mole fraction profiles across the different physical models at the selected heat-bath temperatures. The GB+VS approach successfully reproduces the energy profiles regardless of the existence of the dissociation mechanisms, as shown in Figs. \ref{fig:VS_mix_energy_H2_mole_fraction_ei_wo_diss} and \ref{fig:VS_mix_energy_H2_mole_fraction_ei_w_diss}. This is attributed to the application of GB coarse-graining to the low-lying energy states, which accurately captures the energy transfer over the simulation time. 
In contrast, the VS+CB approach offers slightly improved accuracy compared to the original VS coarse-graining, however, it still significantly overpredicts the initial phase of the internal energy evolution, leading to the faster onset time of the dissociation as shown in Fig. \ref{fig:VS_mix_energy_H2_mole_fraction_H2_molefrac}, similar to the VS coarse-graining. 
This observation implies that the failure of the VS model is primarily due to the inability of the vibration-specific treatment for the low-lying energy levels, resulting in the overprediction of the internal energy relaxation.
In Fig. \ref{fig:VS_mix_energy_H2_mole_fraction_H2_molefrac}, it is important to note that at 5,000 K, the GB+VS model overestimates the amount of dissociation compared to the full StS and the HCG (see Fig. \ref{fig:models_H2_combined_mole_frac}) approaches. This denotes that the dissociation dynamics of the $\text{H}_2\left(\text{X}^1\Sigma_g^+\right)$+$\text{H}\left({}^2\text{S}\right)$ system is favorably governed by the distance from the centrifugal-barrier, rather than the vibration-specific manner, similar to the previous findings for the heavier diatomic molecular species of the air mixtures \cite{jo2022rovibrational,venturi2020data}, and further justifies the HCG approach proposed by the present study.

\subsubsection{\label{sec:model_comp_2D}2-D Axisymmetric Hypersonic Flows}

In this subsection, hypersonic atmospheric entry flows relevant to Uranus planetary exploration \cite{palmer2014aeroheating} are simulated over a 2-D axisymmetric half-sphere configuration with 0.3 m radius to further assess the nonequilibrium model's performance in a realistic flow regime. The freestream flow condition is mostly taken from the previous work by Palmer \textit{et al.} \cite{palmer2014aeroheating}: $p_{\infty}=9.397$ Pa, $T_{\infty}$ = 300 K, $u_{\infty}$ = 22.503 km/s, $X_{\text{H},\infty}$ = 0.001, and $X_{\text{H}_2,\infty}$ = 0.999, respectively, where the subscript $\infty$ denotes the freestream. It is worth noting that the actual $T_{\infty}$ was reported as 128.2 K \cite{palmer2014aeroheating}, however, in the present study, the lower temperature threshold is set to 300 K to avoid numerical instability of the thermodynamic property calculations, which are frequently proportional to $\exp(1/T)$ expression. An isothermal noncatalytic no-slip boundary condition is imposed at 500 K along the body surface. A rectangular type structured mesh topology with $202\times101$ grid size, respectively along the axis and off-axis directions, is used with shock-alignment mesh adaptation. HEGEL \cite{munafo2020computational,alberti2019laser,munafo2024hegel} is used to simulate the steady-state 2-D axisymmetric hypersonic flows with 2nd order accuracy in spatial discretization. It is important to note that in these 2-D axisymmetric CFD calculations, the full StS and the coarse-graining models are not considered due to their high computational cost in the multi-dimensional configuration, therefore, the model comparison is carried out for the existing 2T and the present modified 2T approaches.

\begin{figure}
    \centering
    \subfigure[Flow properties along stagnation-line.]
    {
        \includegraphics[width=0.48\textwidth]{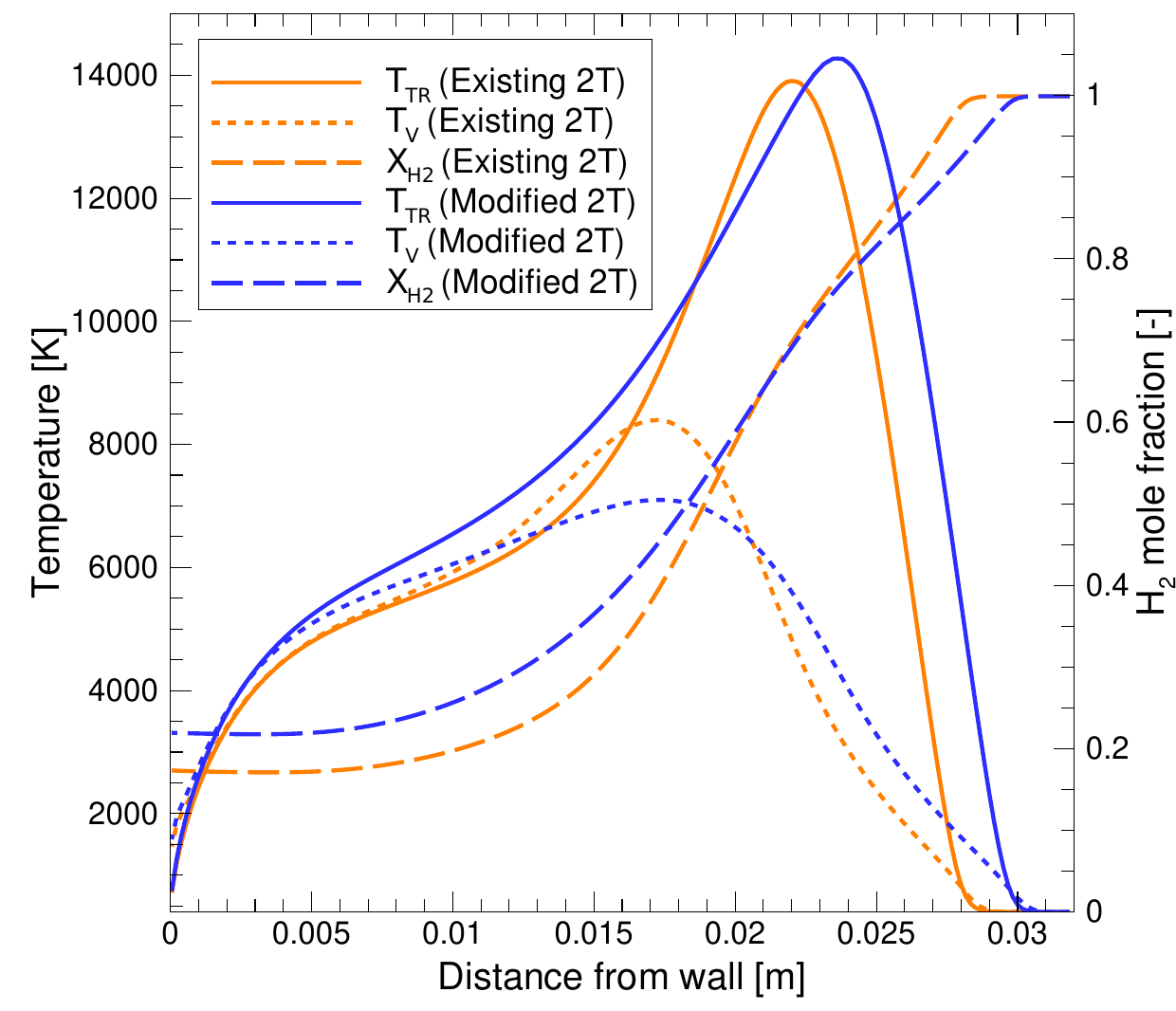}
        \label{fig:Plot_Stag}
    }
    \subfigure[Convective heat flux along wall.]
    {
        \includegraphics[width=0.48\textwidth]{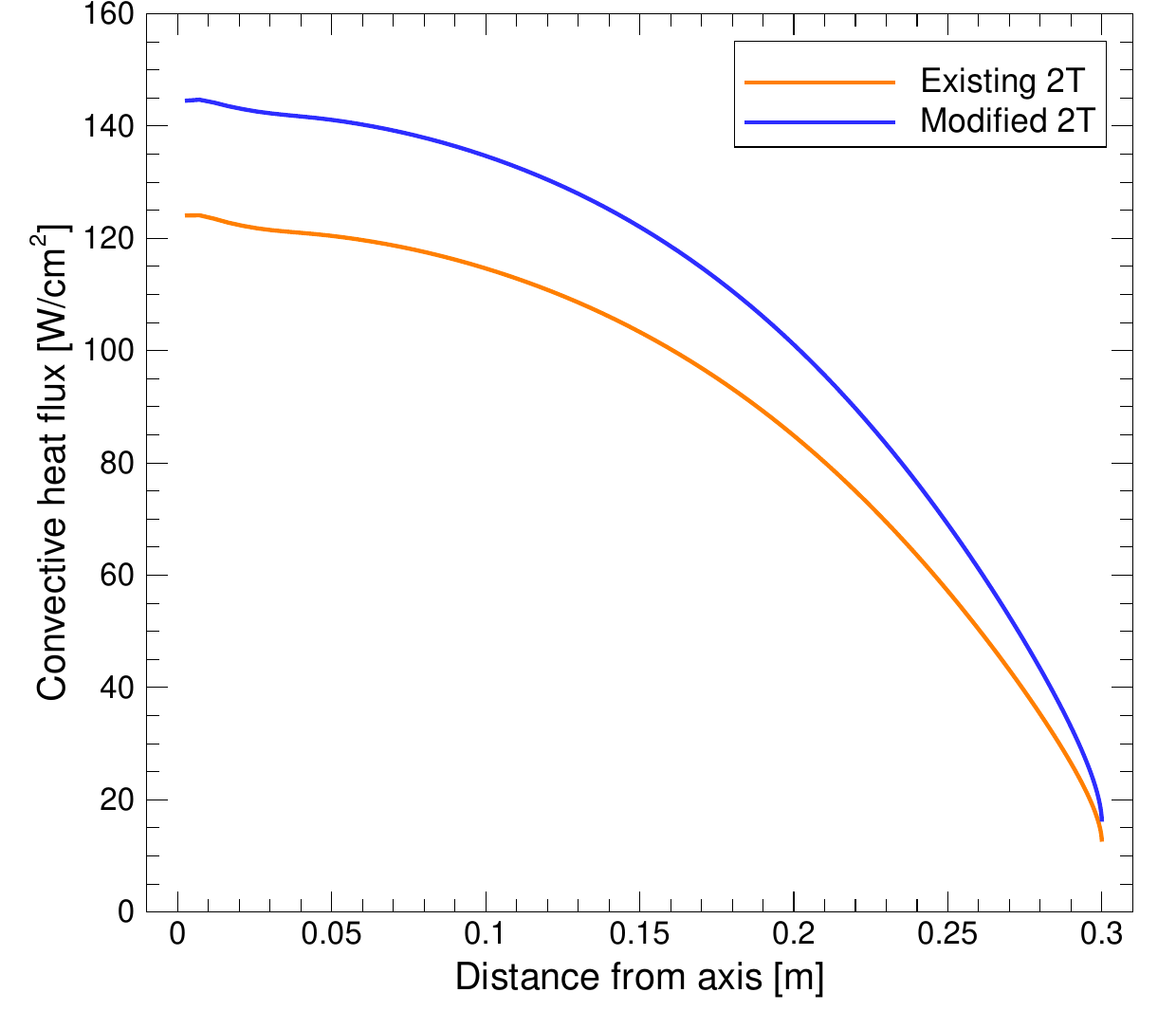}
        \label{fig:Plot_SurfHeatFlux}
    }
  \caption{Comparisons of the modified 2T and existing 2T models for the Uranus hypesonic atmospheric entry flows around the 2-D axisymmetric forebody sphere.}
    \label{fig:Plot_2D}
\end{figure}

Figure \ref{fig:Plot_2D} shows comparisons of the stagnation-line flow properties and the convective heat flux along the wall, predicted by the existing 2T and the modified 2T models. As shown in Fig. \ref{fig:Plot_Stag}, the modified 2T model predicts the larger shock stand-off distance attributed to the slower energy transfer and establishment of the molecular QSS period as observed in the 0-D cases. In addition, the implicit inclusion of the RVT energy transfer through $k_f(T_a)$ in Eq. (\ref{eq:k-fitting}) is also responsible for the increase of the shock stand-off distance of the modified 2T model.
In the existing 2T approach, the well-known overshoot of $T_V$ is observed, whereas it disappears in the modified 2T model. This is mainly due to the updated treatment of the chemistry-vibrational energy coupling through the master equation-informed preferential modeling in Eq.~(\ref{eq:omega_cv}) by defining $C^{DV}$ as the function of temperature. As a result, the peak values of $T_V$ differ by around 1400 K near 1.72 cm away from the stagnation-point. 

From the shock front to the wall, more dissociation occurs in the existing 2T model, leading to less $\text{H}_2$ arriving at the stagnation-point. This is due to the higher dissociation rate coefficient of the existing approach. In terms of aerothermal heating characterization, as shown in Fig. \ref{fig:Plot_SurfHeatFlux}, the modified 2T model predicts a higher convective heat flux profile than the existing 2T approach, with the maximum discrepancy of around 16.5\% at the stagnation-point. This is attributed to the lower degree of chemical dissociation (\emph{i.e.}, endothermic process) predicted by the modified model, leading to a higher residual sensible enthalpy that can be conducted to the wall in the absence of catalytic surface effects. The result shown in Fig. \ref{fig:Plot_SurfHeatFlux} demonstrates the importance of accurate prediction of the nonequilibrium chemical-kinetic process of the $\text{H}_2$+H system for the aerothermal characterization relevant to the Uranus planetary entry probe.

\section{\label{sec:5}Conclusions}

The present study performed detailed rovibrational-specific QCT and master equation analyses for the $\text{H}_2\left(\text{X}^1\Sigma_g^+\right)$+$\text{H}\left({}^2\text{S}\right)$ system using the most recent \emph{ab-initio} BH PES. The full set of rovibrational StS kinetic database, covering a wide range of temperatures, enabled the characterization of the energy transfer and dissociation mechanisms, including the system's unique relaxation behavior of the vibrational energy mode during the early stage. Then, the two-different reduced-order models, the modified two-temperature (2T) and the hybrid coarse-graining (HCG), were proposed, informed by the rovibrational StS master equation analysis. The modified 2T model was designed by incorporating the updated nonequilibrium chemical-kinetic parameters, including the vibrational relaxation time with the high-temperature correction, the quasi-steady-state (QSS) dissociation rate coefficient that implicitly incorporates the rotational-vibrational-translational (RVT) energy transfer, and the average vibrational energy loss ratio as a function of temperature, enabling preferential treatment of the nonequilibrium dissociation. The HCG model was proposed by combining a graph-based coarse-graining optimized by maximum likelihood estimation with a dissociation-favorable coarse-graining based on the distance from the centrifugal-barrier, enabling simulations over a wide range of temperatures, unlike previous studies limited to a specific temperature.

The detailed model comparisons in 0-D isothermal and isochoric heat bath calculations and in 2-D axisymmetric hypersonic flow simulations demonstrate the significantly improved accuracy of the reduced-order models proposed in the present study. Both the modified 2T and the HCG models more accurately predict the nonequilibrium relaxation of the internal energy, vibrational temperature, and species concentration than the existing coarse-graining approaches, such as the energy-based (EB) and vibration-specific (VS) models, and the existing 2T approach. The failure of conventional VS coarse-graining is quantitatively characterized for the first time, revealing the importance of advanced treatment of energy transfers among low-lying energy states. Finally, the modified 2T and existing 2T models were compared by simulating 2-D axisymmetric hypersonic flows relevant to Uranus planetary entry. By modeling the RVT energy transfer and the advanced treatment of the preferential dissociation of the present modified 2T model, the larger shock stand-off distance and the higher convective heat flux by 16.5\% were predicted compared to the existing 2T model found in the literature, which demonstrates the importance of accurate modeling of the nonequilibrium chemical-kinetics of the $\text{H}_2\left(\text{X}^1\Sigma_g^+\right)$+$\text{H}\left({}^2\text{S}\right)$ system for appropriate aerothermal characterization of the relevant hypersonic flows.

\section*{Acknowledgments}
The work was partially supported by AFOSR Grant No. FA9550-25-1-0119 with Dr. Amanda Chou as Program Manager (to S.M.J.). The authors thank Dr. A. Munafò, Dr. M. Panesi (University of California, Irvine), and Dr. S. Venturi (Atomic Machines) for providing access to the \textsc{plato}, \textsc{hegel}, and \textsc{CoarseAIR} software.

\bibliography{ref}

\end{document}